\documentclass[12pt]{article}
\usepackage{mathtools,amssymb,amsfonts,graphicx,tensor}

\usepackage{hyperref}
\usepackage[nosort]{cite}
\usepackage{verbatim}
\usepackage{multicol,color,xcolor,longtable}
\usepackage{hyperref}
\usepackage{cleveref}

\usepackage{amsmath} 
\usepackage{cleveref}
\crefformat{equation}{(#2#1#3)}
\crefrangeformat{equation}{(#3#1#4--#5#2#6)}
\crefmultiformat{equation}{(#2#1#3)}{ and~(#2#1#3)}{, (#2#1#3)}{ and~(#2#1#3)}

\definecolor{darkred}{rgb}{0.65,0.15,0}
\definecolor{darkgreen}{rgb}{.05,.5,.05}
\hypersetup{pdfborder={0 0 0},colorlinks=true,urlcolor=darkred,citecolor=blue,linkcolor=darkred,linktocpage=true}

\usepackage{soul}
\setulcolor{green}

\newcommand{\cV}{\mathcal{V}}

\makeatletter

\@addtoreset{equation}{section}
\makeatother

\newcommand{\be}{\begin{equation}}
\newcommand{\ee}{\end{equation}}

\begin{document}

\mbox{}

\vspace{20mm}

\begin{center}
{\LARGE \sc  The auxiliary-deformed Breitenlohner-Maison model: duality frames and higher-dimensional origin}    
\\[13mm]

{\large
Daniele Bielli${}^{1,2}$ and Mattia Ces\`aro${}^{3}$ }

\vspace{8mm}
${}^{1}${\it 
Asia Pacific Center for Theoretical Physics (APCTP), \\ 
Postech, Pohang 37673, 
Korea}
\\~\\
${}^{2}${\it 
High Energy Physics Research Unit, Faculty of Science \\ 
Chulalongkorn University, Bangkok 10330, Thailand
}
\\~\\
${}^3${\it Max-Planck-Institut f\"{u}r Gravitationsphysik (Albert-Einstein-Institut)\\
Am M\"{u}hlenberg 1, DE-14476 Potsdam, Germany}

\vspace{3mm}
~\\
\texttt{d.bielli4@gmail.com,
mattia.cesaro@aei.mpg.de
}\\
\vspace{3mm}

\end{center}






\begin{center}

\hrule

\vspace{6mm}

\begin{tabular}{p{14cm}}
{\small%
The two-dimensional Breitenlohner-Maison (BM) model is a classically integrable subsector of $D=4$ general relativity endowed with two commuting Killing isometries, obtained via Kaluza-Klein reduction to $D=2$. Integrable deformations of such a theory have recently been constructed via auxiliary fields in the so-called $\nu$-frame. In this work we first extend this point of view by deriving the complementary auxiliary field perspective known as $\mu$-frame, and then explicitly construct the uplift to $D=4$ of both descriptions, relying on an ansatz inspired by duality-invariant Lagrangian formulations of Einstein theory. The resulting four-dimensional deformed model thus obtained is a higher-derivative theory which lacks manifest diffeomorphism invariance in both frames. We comment on possible resolutions of this puzzling feature and on the physical interpretation of the model in $D=4$. }
\end{tabular}
\vspace{5mm}
\hrule
\end{center}

\hypersetup{pageanchor=false}
\thispagestyle{empty}

\newpage
\hypersetup{pageanchor=true}
\setcounter{page}{1}


\tableofcontents
\newpage
\section{Introduction}
The space of solutions of Einstein equations in $D=4$ equipped with two commuting Killing isometries is well studied, and stands out among other classes of solutions because of its remarkable features, among which the complete (classical) integrability of the field equations \cite{Belinsky:1971nt,Maison:1978es}. In particular, the theory can be phrased in terms of an effective two-dimensional model featuring a rich algebraic structure, obtained via a Kaluza-Klein (KK) reduction along the two isometries from $D=4$ to $D=2$. 
The Lagrangian can be cast in the form of a coset sigma-model coupled to a dilaton and to a $D=2$ gravitational field. Its complete integrability does not come as a coincidence, but it is rather deeply connected to the presence of a hidden on-shell, infinite-dimensional duality group of the affine Kac-Moody type, $\widehat{\text{SL(2)}}$, known as the Geroch group \cite{Geroch:1970nt, Geroch:1972yt}, whose action maps solutions of the equations of motion into other solutions. The presence of the central extension underlying this infinite dimensional structure was first foreseen in \cite{Julia:1980gr, Julia:1981wc}, but its precise action on the physical fields was only thoroughly elucidated by Breitenlohner and Maison in \cite{Breitenlohner:1986um}. In the rest of this work, we will thus refer to this $D=2$ model as the``Breitenlohner-Maison" (BM) model.  

The BM model furnishes an interesting example lying at the interface between integrable systems, duality symmetries and gravitational theories. A consistent amount of work has been devoted in the literature to the exploration of this two-dimensional model, in particular to its quantisation~\cite{Korotkin:1994au, Korotkin:1996vi, Korotkin:1996au, Korotkin:1996fx, Korotkin:1997fi}, and to its generalisations to supergravity theories~\cite{Nicolai:1987kz,Nicolai:1988jb} and to exceptional field theory~\cite{Bossard:2018utw, Bossard:2021jix, Bossard:2023wgg}. Recently in~\cite{Cole:2024skp}, the model has also been shown to arise from the framework of $D=4$ Chern-Simons theory, an approach that has been proven to underlie various integrable field theories in $D=2$ ~\cite{Costello:2019tri,Vicedo:2019dej,Delduc:2019whp} (and on which we will comment more in the concluding section of this work). 

 Integrable theories, both at the classical and quantum level, form an intriguing and at the same time useful class of models. Despite being interacting and exhibiting highly non-trivial dynamics, it is often possible to leverage the rich underlying symmetry structure and to apply exact analytic techniques in order to compute physical observables of interest. This represents a quite uncommon feature in comparison to general field theories. Exact computations are indeed typically beyond reach, and for this reason integrable theories are singled out as very appealing theoretical laboratories where to test and learn about realistic physical phenomena as well as mathematical methods \cite{Babelon:2003qtg,Dorey:2019gkd,Bombardelli:2016rwb,Beisert:2010jr}. 

A new family of classically integrable deformations for $D=2$ sigma models, inspired by non-linear reformulations of $D=4$ Maxwell's theory \cite{Ivanov:2001ec ,Ivanov:2002ab}, and the observations in \cite{Borsato:2022tmu,Ferko:2023wyi}, has been lately introduced. This was originally constructed for the Principal Chiral Model in the seminal paper \cite{Ferko:2024ali} -- soon also recovered from the $D=4$ Chern-Simons perspective \cite{Fukushima:2024nxm} -- and relies on the introduction of auxiliary fields (in the sense that their field equations are purely algebraic) into the Lagrangian of an integrable ``seed" theory. These auxiliaries couple in a well prescribed way to the physical fields, and the Lagrangian is modified by the further addition of a deforming function of a certain invariant combination of auxiliary fields only. These deformations generate continuous, infinite families of integrable theories, encompassing the well-known $T\bar{T}$ and ``root-$T\bar{T}$" ~\cite{Zamolodchikov:2004ce,Cavaglia:2016oda,Ferko:2022cix,Jiang:2019epa,He:2025ppz} and extending them to deformations by arbitrary function of the stress-tensor.

Deformations by higher-spin conserved currents of the Smirnov-Zamolodchikov type \cite{Smirnov:2016lqw} have also been realised using this framework \cite{Bielli:2024ach,Bielli:2025uiv}, which has furthermore been applied to various other classes of $D=2$ theories \cite{Bielli:2024khq,Bielli:2024fnp,Cesaro:2024ipq, Bielli:2024oif, Cesaro:2025msv}, revealing interesting underlying mathematical structures \cite{Ferko:2025bhv,Baglioni:2025tsc,Bielli:2026ggs}. Particularly relevant to this work are the results in \cite{Baglioni:2025tsc}, where it was shown that such auxiliary field deformations are amenable to a different duality frame description, known as the $\mu$-frame \cite{Ivanov:2003uj}, comparatively less studied in the literature, with respect to the original formulation in the so-called $\nu$-frame. The two perspectives, related via a Legendre transform on a subset of interaction functions, complement each other without fully overlapping. Hence, their parallel exploration constitutes in itself an interesting research endeavour, contributing to a deeper understanding of integrable field theories.

The focus of the present work is on \cite{Cesaro:2025msv}. Here, the authors constructed the auxiliary field setup relevant to the BM model, thus leading to an infinite family of its deformations preserving both the integrability of the deformed field equations and the presence of the Geroch group. The model presents some similarities and at the same time some subtle differences  (among which, for example, the presence of a physical field in the deformation function) in comparison the other auxiliary field deformed theories, and its derivation led to further interesting issues remained so-far unresolved. More specifically, a natural question that came associated to the $D=2$ construction is whether this deformed lower-dimensional model admits an uplift back to $D=4$. This would be to some extent surprising: despite the appearance of the Geroch group in the deformed $D=2$ model, it was conjectured in \cite{Cesaro:2024ipq} that such an uplift would entail higher-derivative corrections to Einstein theory, and the presence of the latter is usually known to break (in part or completely) the hidden duality groups upon KK-reduction to lower dimensions~\cite{Michel:2007vh, Lambert:2006he, Lambert:2006ny, Bao:2007er} (with the exception of trivial field redefinitions, see \cite{Colonnello:2007qy}). 

The aim of this paper is twofold. On the one hand, we engineer the $\mu$-frame description of the BM model, thus extending the results of \cite{Baglioni:2025tsc} to the theory in question. This formulation of the deformed theory is expressed in terms of a single scalar auxiliary field, rather than multiple Lie-algebra valued ones, and can simplify the proof of certain integrable properties. On the other hand, we address the question opened in \cite{Cesaro:2025msv}, and construct the explicit uplift to $D=4$ of the deformed model in both the $\nu$- and $\mu$-frames: our results confirm, as expected for this class of deformations, the higher-derivative nature (albeit a very peculiar one) of the theory also in higher-dimensions. The higher-derivative terms appearing in $D=4$ are not the most general ones allowed, since $n$-th order derivatives of the metric never appear. This might hint at a possible mechanism by which the affine symmetry could be preserved upon reduction to $D=2$.   

The work is organised as follows. In section \ref{sec:Review}, we review the $D=2$ auxiliary field deformation of the PCM in both $\nu$- and $\mu$-frames (in section \ref{subsec:framesforPCM}), together with the $\nu$-frame deformation of the BM model \cite{Cesaro:2025msv} (in section \ref{subsec:nuframeBM}). In section \ref{sec:muframeBM}, following the strategy employed in \cite{Baglioni:2025tsc}, we proceed to derive the $\mu$-frame formulation of the deformed BM model. Section \ref{sec:uplifts} is devoted to the $D=4$ uplift of the deformed BM model in both the $\nu$- and $\mu$-frames and to the exploration of its symmetries (or rather lack thereof), in section \ref{sec:symmetries}. We conclude by commenting on the puzzling features of the uplifts and by pointing at future research directions in section \ref{sec:conclusions}.
More technical computations, as well as the KK-reduction of the geometrical objects needed in the body of the paper, are collected in several appendices.

\section{Review of $D=2$ auxiliary field deformations}\label{sec:Review}
\subsection{Principal chiral models in the $\nu$ and $\mu$ frames}\label{subsec:framesforPCM}
The Principal Chiral Model (PCM) represents one of the most studied examples of two-dimensional sigma models. The fundamental fields of the theory are coordinates on a Lie group G and their interaction can be understood as the metric on such a background. The action can be written in terms of the pull-back to the worldsheet $\Sigma$ of the Lie-algebra-valued Maurer-Cartan form $j:=\mathrm{g}^{-1}\mathrm{d}\mathrm{g} \in \mathfrak{g}=$Lie(G): 
\begin{equation}\label{PCMundeformed}
S_{\text{PCM}}=
-\tfrac{1}{2}\int_{\Sigma} \mathrm{d}x^{+}\mathrm{d}x^{-} \mathrm{Tr}(j_{+}j_{-}) \,\, .
\end{equation}
The symbol $\mathrm{Tr}$ denotes trace over the Lie algebra generators in the adjoint representation and we are using lightcone coordinates
\begin{equation}\label{eq:lightcone-coordinates-def}
x^{\pm}=\tfrac{1}{2}(t\pm x)
\quad \text{such that} \quad
\eta_{\alpha\beta}=
\begin{pmatrix}
0 & -2
\\
-2 & 0
\end{pmatrix}
\quad \text{and} \quad 
\eta^{\alpha\beta}=
\begin{pmatrix}
0 & -\tfrac{1}{2}
\\
-\tfrac{1}{2} & 0
\end{pmatrix} \,\, .
\end{equation}
The Maurer-Cartan form plays in this context the role of a vielbein on the background, with the trace on the Lie-algebra being the tangent space metric. The fact that PCMs can be fully encoded in this object, has contributed to making these theories ideal arenas to test and gain a better understanding of ideas related to two-dimensional integrability. For instance, they have been exploited as "seed" theories for the construction of various classes of integrable deformations. Notable examples in this sense include deformations by a Wess-Zumino term \cite{Wess:1971yu,Novikov:1982ei,Witten:1983ar}, by operators on the underlying Lie algebra built out of $\mathcal{R}$-matrices \cite{Klimcik:2002zj,Klimcik:2008eq,Klimcik:2014bta}, or even theories interpolating between the so-called non-Abelian T-dual model of the PCM and WZW models \cite{Sfetsos:2013wia}. See also the reviews \cite{Zarembo:2017muf,Lacroix:2018njs,Hoare:2021dix,Borsato:2023dis} for pedagogical treatments and a more detailed list of references, as well as \cite{Bahadori:2026iig} for another interesting recently proposed integrable deformation.\\
\indent
The class of deformations we shall be concerned with in this work are also of recent construction and rely on the idea of coupling the seed theory to a set of auxiliary fields exhibiting certain types of interactions. Inspired by the work of Ivanov and Zupnik in the context of $D=4$ duality-invariant theories of non-linear electrodynamics \cite{Ivanov:2001ec ,Ivanov:2002ab,Ivanov:2003uj}, and the observations in \cite{Borsato:2022tmu,Ferko:2023wyi}, these deformations enjoy two formulations, or frames, named after the auxiliary variables characterising each of them. Here and in the rest of the work we  use the symbol ``$\doteq$" for equalities valid when the auxiliary field equations are satisfied, and the symbol ``$\approx$" for equalities holding on the physical field equations. These are further combined into ``$\dot{\approx}$" for fully on-shell equalities.

\paragraph{$\nu$-frame.} In this description, constructed in \cite{Ferko:2024ali}, the PCM is deformed by introducing Lie-algebra-valued auxiliary fields $v$ with simple coupling to the Maurer-Cartan form and characterised by a self-interaction function that does not involve the physical fields\footnote{Notice that, throughout this work, and differently from previous literature, we denote the interaction function by $\mathcal{E}$, rather than $E$, which we reserve for the vierbein determinant.}
\begin{equation}\label{nu-frame-PCM}
\mathcal{L}_{\text{PCM}}^{(\nu)}=\tfrac{1}{2}\mathrm{Tr}(j_{+}j_{-})+\mathrm{Tr}(j_{+}v_{-}+v_{+}j_{-})+\mathrm{Tr}(v_{+}v_{-})+\mathcal{E}(v) \,\, .
\end{equation}
The EOM for the auxiliary fields read
\begin{equation}
v_{\pm} \doteq -j_{\pm}-\Delta_{\pm} \,\, ,
\qquad \text{with} \qquad 
\Delta_{\pm}:=\frac{\delta \mathcal{E}(v)}{\delta v_{\mp}} \,\, ,
\end{equation}
and highlight how for trivial self-interactions, namely $\mathcal{E}=0$, they can be easily integrated out recovering the undeformed PCM. Non-trivial choices of interaction functions have been shown to induce deformations by functions of the stress-tensor \cite{Ferko:2024ali} and of higher-spin conserved currents \cite{Bielli:2024ach}, according to the specific dependence of $\mathcal{E}$ on $v$, and compatibility with Lax integrability is respected provided that\footnote{$N$ in equation \eqref{Delta-v-commutator} being a large enough integer to guarantee having a complete set of algebraic structures describing completely symmetric invariant tensor.} \cite{Bielli:2025uiv} 
\begin{equation}\label{Delta-v-commutator}
[\Delta_{\pm},v_{\mp}]\doteq0 
\quad \,\, \text{for} \quad \,\,
\mathcal{E}(v)\!:=\!\mathcal{E}(\nu_{-N},...,\nu_{-2},\nu_{+2},...,\nu_{+N}),
\quad \,\, \text{with} \quad \,\, 
\nu_{\pm k}:= \mathrm{Tr}(v_{\pm}^{k})  .
\end{equation}
In this work we will focus on the case of stress-tensor deformations, encoded in
\begin{equation}
\mathcal{E}(v):=\mathcal{E}(\nu) 
\qquad \text{with} \qquad
\nu:=\sqrt{\mathrm{Tr}(v_{+}^2)\mathrm{Tr}(v_{-}^2)} \,\, ,
\end{equation}
using a notation which is slightly different from the original paper, but turns out to be convenient for the below alternative formulation, as also highlighted around \eqref{eq:sqrtnu}.

\paragraph{$\mu$-frame.} This description of the deformed PCM was constructed in \cite{Baglioni:2025tsc}
\begin{equation}
\mathcal{L}_{\text{PCM}}^{(\mu)}=\frac{2\mu \sqrt{\mathrm{Tr}(j_{+}j_{+})\mathrm{Tr}(j_{-}j_{-})}-(1+\mu^2)\mathrm{Tr}(j_{+}j_{-})}{2(1-\mu^2)}+\mathcal{H}(\mu) \,\, ,
\end{equation}
by performing a Legendre-transform of \eqref{nu-frame-PCM} with respect to the variable $\nu$
\begin{equation}\label{PCM-Legendre-transform}
\mathcal{H}(\mu):=\mathcal{E}(\nu)-\nu \mathcal{E}_\nu(\nu),
\qquad \text{with} \qquad 
\mu:=\mathcal{E}_\nu(\nu) \,\, .
\end{equation}
Here, we have denoted by $\mathcal{E}_\nu(\nu)$ the derivative of $\mathcal{E}(\nu)$ with respect to $\nu$. One of the key differences with the $\nu$-frame lies in the structure of the auxiliary variables which induce the deformation: upon Legendre-transforming, the Lie-algebra-valued fields $v$ are traded for a single scalar $\mu$, whose equation of motion can be more easily computed 
\begin{equation}
\frac{(1+\mu^2) \sqrt{\mathrm{Tr}(j_{+}j_{+})\mathrm{Tr}(j_{-}j_{-})}-2\mu\mathrm{Tr}(j_{+}j_{-})}{(1-\mu^2)^2}\doteq - \mathcal{H}_\mu(\mu) \,\, ,
\end{equation}
where $\mathcal{H}_\mu(\mu)\equiv \frac{\partial\mathcal{H}(\mu)}{\partial\mu}$. A second difference can be observed in the recovery of the undeformed theory. Contrarily to the previous case, by integrating out $\mu$ in the presence of a trivial interaction function leads to a non-analytic Lagrangian, while the PCM is instead recovered, still after setting $\mathcal{H}=0$, in the limit where $\mu \rightarrow 0$. This can in fact be also regarded, at the level of Legendre-transform \eqref{PCM-Legendre-transform}, as the limit where the $\nu$-frame interaction function $\mathcal{E}$ is trivialised. It should also be mentioned that convexity requirements for the invertibility of the Legendre transform make the two frames equivalent only on a restricted subset of the allowed interaction functions. Either description could indeed more generally include interesting deformations unreachable from the other one, making it important to gain a better understanding of both frames, each of which seems indeed to exhibit its own advantages and disadvantages. The already noted change in the structure of the auxiliary variables is for example accompanied, as observed in \cite{Baglioni:2025tsc} for the case of $T\bar{T}$ and root-$T\bar{T}$, by a swap in the complexities of the Lagrangian and interaction function, which are respectively simple and complicated in the $\nu$-frame, while exhibiting the opposite trend in the $\mu$-frame.

\subsection{The Breitenlohner-Maison model in the $\nu$-frame}\label{subsec:nuframeBM}
The model under consideration in the present work is
 the dimensional reduction of $D=4$ general relativity along two commuting space-like Killing isometries (for comprehensive treatments, see, for example, \cite{Nicolai:1991tt,Nicolai:1996pd}), and the auxiliary field deformations thereof \cite{Cesaro:2025msv}.
 Details on the general structure of $\mathbb{Z}_2$ coset geometries, instrumental in the following construction, are summarised in appendix \ref{secion:appendixcoset}.

We work with conventions $\eta_{AB}=\text{diag}(-,+,+,+)$ and, when useful, adopt lightcone coordinates \eqref{eq:lightcone-coordinates-def} in two dimensions.
Starting with the $D=4$ Einstein-Hilbert Lagrangian,
\begin{equation}\label{eq:EH}
 \mathcal{L}^{(4)}_{\text{EH}}=\frac{1}{2}E\,R(E)_{(4)}\, ,   
\end{equation}
with $E$ the vierbein determinant and $R(E)_{(4)}$ the Ricci scalar, the KK reduction to $D=2$ furnishes a lower-dimensional model consisting of $D=2$ gravity coupled to a coset sigma model. The $2D$ theory is endowed with a global on-shell hidden symmetry, enhanced to the central affine loop extension of SL$(2)$, the Geroch group $\widehat{\text{SL}(2)}$  \cite{Geroch:1972yt,Geroch:1970nt}\footnote{One way $\widehat{\text{SL}(2)}$ can be seen to arise is to try to consistently implement the simultaneous action on the fields of the two different SL$(2)$'s arising from respectively reducing the $D=4$ theory directly on $T^2$ (the Matzner-Misner one), or dualising first the vector field in $D=3$ and then reducing to $D=2$ (the Ehlers one). This requires the introduction of an infinite tower of on-shell conserved dual potentials.}. This infinite-dimensional symmetry is ultimately responsible for the integrability of the model \cite{Breitenlohner:1986um} and its action maps solutions of the equations of motion to other solutions.

For definiteness, we will consider the two isometries to be space-like, and the dimensional reduction to be the one on $T^2$, implemented through the choice of vierbein (the so-called \textit{Matzner-Misner} frame) 
\begin{equation} \label{eq:vierbein1}
    E_M{}^A=\begin{pmatrix}
        e_\mu{}^\alpha & B_\mu{}^m e_m{}^a \\
        0 &e_m{}^a
    \end{pmatrix}=\begin{pmatrix}
        e_\mu{}^\alpha &0\\
        0& \rho^{\frac{1}{2}}\mathcal{V}_m{}^a
    \end{pmatrix}\, .
\end{equation}
Here, we have split the curved spacetime indices as $M=(\mu, m)$, and the flat Minkowski ones as $A=(\alpha, a)$. $\mathcal{V}_m{}^a$ is a SL(2)/SO(2) representative, transforming as in \eqref{eq:costrm}, with global $g\in \text{SL(2)}$ and local (in space-time) $k(x) \in \text{SO(2)}$. $e_\mu{}^{\alpha}$ is the external zweibein and the Kaluza-Klein vector $B_\mu{}^m$ carries no physical degree of freedom anymore in $D=2$, and acts as an auxiliary field, thus to be effectively set to a constant (0, for asymptotically Minkowski solutions). The lower-right block can be parameterised by its own determinant (measuring the size of $T^2$), $\rho\equiv \text{det} e_m{}^a$, and an SL$(2,\mathbb{R})$ matrix $\mathcal{V}_m{}^a$. Inserting directly \eqref{eq:vierbein1} (with $B_\mu{}^m=0$) into \eqref{eq:EH}, leads to the effective two-dimensional Lagrangian
\begin{equation}
  {\mathcal{L}}^{(2)}_{\text{EH}}= \frac{1}{2}e_{(2)}\rho R_{(2)}-\frac{1}{2}e_{(2)} \rho \text{Tr}(P_\mu P^\mu)+\frac{1}{4} e_{(2)}\rho h^{\mu\nu}\partial_\mu \text{ln}\rho\partial_\nu\text{ln}\rho\, ,
\end{equation}
where $P_\mu$ is the $\text{SL}(2,\mathbb{R})/\text{SO}(2)$  coset current as defined in \eqref{eq:QP}.

By choosing the conformal gauge for the $D=2$ metric, namely
\begin{equation}
 h_{\mu\nu}= e^{2 \sigma} \eta_{\mu\nu},  
\end{equation}
and rescaling $\sigma$ as
\begin{equation}
  \sigma \rightarrow \sigma+\frac{1}{4}\text{ln}\rho\ ,  
\end{equation}
the Lagrangian can be cast in the remarkably simple form
\begin{align}\label{eq:LagMM}
{\mathcal{L}}^{(2)}_{\text{EH}}= \partial_\mu\rho\partial^\mu\sigma- \frac{1}{2}\rho \text{Tr}(P_\mu P^\mu)\, .
\end{align}
Its equations of motion\footnote{Notice that  equation \eqref{eq:boxsigma0} is not fundamental, and it is automatically implied if \eqref{eq:DP0}, \eqref{eq:boxrho0}, \eqref{eq:Vir2} and \eqref{eq:Bianchis} are satisfied.},
\begin{align}
 &D_\mu(\rho P^\mu)=0\,, \label{eq:DP0}\\
 &\Box\sigma+\frac{1}{2} \text{Tr}(P_\mu P^\mu)=0\,,\label{eq:boxsigma0}\\
 &\Box \rho=0\,\label{eq:boxrho0},
\end{align}
 must be supplemented  with the associated Virasoro constraints, 
\begin{align}
&\partial_\pm\rho\partial_\pm\sigma-\frac{1}{2}\partial_\pm\partial_\pm\rho-\frac{1}{2}\rho \text{Tr}(P_\pm P_\pm)=0\,,\label{eq:Vir2}    
\end{align}
written here in lightcone coordinates and formally obtained by varying the action with respect to the unimodular degrees of freedom before gauge-fixing to the conformal gauge.

The family of integrable deformations of \eqref{eq:LagMM} via appropriate coupling to the auxiliary fields has been obtained in the $\nu$-frame in \cite{Cesaro:2025msv}. The relevant Lagrangian is
\begin{equation}\label{eq:Ldef}
\begin{aligned}\mathcal{L}^{(2,\;\nu)}=&-\partial_\mu \rho\partial^\mu\sigma -2(\chi_{1\,\mu} \partial^\mu \rho+\chi_{2\,\mu}\partial^\mu\sigma+\chi_1^\mu\chi_{2\,\mu})
\\
& +\rho\text{Tr}\left[\tfrac{1}{2}(P_\mu P^\mu)+2P_\mu v^\mu+v_\mu v^\mu\right]+\mathcal{E}(\nu_2)\, ,
\end{aligned}
\end{equation}
where $\left(\chi_{1\,\mu}, \chi_{2\,\mu}, v_\mu\right)$ are a set of auxiliary fields (i.e. their equations of motion are algebraic), with $v_\mu \in \mathfrak{sl}_2\ominus\mathfrak{so}_2$. $\mathcal{E}(\nu_2)$ is an arbitrary function of the parameter $\nu_2$,  \footnote{We stress the difference between $\nu_2$ here defined and its redefinition via a power of $\frac{1}{2}$ in the following sections, instrumental in making contact with the related $\mu$-frame. For the sake of clarity we will try to keep these distinctions clear by adopting the notation $\nu_2$ for the object in \eqref{eq:nu}, $\nu_4$ for its 4-dimensional uplift, and $\nu$ for its square-rooted version.}
\begin{equation}\label{eq:nu}
\nu_2=(\eta^{\alpha\beta}\eta^{\gamma\sigma}+\epsilon^{\alpha\beta}\epsilon^{\gamma\sigma})\left(\chi_{1\,\alpha}\chi_{2\, \gamma}-\frac{\rho}{2}\text{Tr}(v_\alpha v_\gamma)\right)\left(\chi_{1\,\beta}\chi_{2\, \sigma}-\frac{\rho}{2}\text{Tr}(v_\beta v_\sigma)\right)\, .
\end{equation}
Defining 
\be\label{eq:J}
\mathcal{P}_\mu=-(P_\mu+2v_\mu)\, ,\quad \mathcal{R}_{\, \mu}=-(\partial_\mu\rho+2\chi_{2\,\mu})\, ,\quad \mathcal{S}_{\, \mu}=-(\partial_\mu\sigma+2\chi_{1\,\mu})\, ,
\ee
the corresponding Euler-Lagrange equations read 
\begin{align}
\partial_\mu\mathcal{R}_{}^\mu&=0\, ,\label{eq:Boxrho=smth}\\
D_\mu(\rho\mathcal{P}^\mu)&=0\, ,\label{eq:dynamicaleom} \\
\partial_\mu \mathcal{S}^\mu&=\text{Tr}\left[\tfrac{1}{2}(P_\mu P^\mu)+2P_\mu v^\mu+v_\mu v^\mu\right]+K^{\mu\nu}(\chi_{1,2},v,\rho)\,\text{Tr}(v_\mu v_\nu) \, ,\label{eq:Boxsigma+smth} \\
P_\mu&\doteq-v_\mu+ K_\mu{}^\nu(\chi_{1,2},v,\rho) v_\nu,\label{eq:algebraicP}\\
\partial_\mu\sigma&\doteq-\chi_{1\,\mu}+ K_\mu{}^\nu(\chi_{1,2},v,\rho)\chi_{1\,\nu}\, ,\label{eq:algebraicsigma}\\
\partial_\mu\rho &\doteq-\chi_{2\,\mu}+K_\mu{}^\nu(\chi_{1,2},v,\rho)\chi_{2\,\nu}\, ,\label{eq:algebraicrho}
\end{align}
where
\begin{equation}
K_\mu{}^\lambda(\chi_{1,2},v,\rho)=\mathcal{E}_{\nu_2}(\delta_\mu{}^\alpha\eta^{\gamma\lambda}-\eta_{\mu\rho}\epsilon^{\rho\alpha}\epsilon^{\gamma\lambda})\left(\chi_{1\,\alpha}\chi_{2\,\gamma}-\frac{\rho}{2}\text{Tr}(v_\alpha v_\gamma)\right)
\end{equation}
is symmetric in the indices $(\mu,\lambda)$ when both are raised or lowered via the flat metric. We also introduced the notation $\mathcal{E}_{\nu_2}\equiv\frac{\partial \mathcal{E}(\nu_2)}{\partial \nu_2}$, which will be used throughout the manuscript. 

The equations of motion \cref{eq:Boxrho=smth,eq:dynamicaleom,eq:Boxsigma+smth,eq:algebraicP,eq:algebraicrho,eq:algebraicsigma} must be accompanied by the relevant deformed Virasoro constraints, 
\begin{equation}\label{eq:deformedVirasoro}
  \tilde{\mathcal{S}}_\pm=-\frac{1}{2}\rho\,\text{Tr}\left(\frac{v_\pm v_\pm}{\chi_{2\,\pm}}+ \frac{K_\pm{}^\mp v_\mp v_\mp}{\chi_{2\,\mp}}\right)\, .
\end{equation}
Here, $\tilde{\mathcal{S}}_\pm$ is a modified version of $\mathcal{S}_\pm$ (namely the lightcone components of $\mathcal{S}_\mu$ defined in \eqref{eq:J}), satisfying $\partial^\mu\tilde{\mathcal{S}}_\mu=\partial^\mu\mathcal{S}_\mu$. Therefore,
\begin{equation}
    \tilde{\mathcal{S}}_\pm=-\partial_\pm\tilde{\sigma}-2\chi_{1\, \pm},
\end{equation}
with $\tilde{\sigma}$ a specifically rescaled version of $\sigma$, whose explicit form is not relevant, and such that, for $K_\pm{}^\mp=0$, 
\begin{equation}
\tilde{\sigma}\Big\lvert_{K_\pm{}^\mp=0}=\sigma+ \frac{1}{2}\text{ln}(\partial_+\rho\partial_-\rho).
\end{equation}

\section{Construction of the $\mu$-frame for the BM model}\label{sec:muframeBM}
 Following the strategy employed in  \cite{Baglioni:2025tsc}, in this section we construct the $\mu$-frame description of the theory obtained in \cite{Cesaro:2025msv}. We start from the auxiliary field deformed Lagrangian in the $\nu$-frame \eqref{eq:Ldef}, written in lightcone coordinates using \eqref{eq:lightcone-coordinates-def}
\begin{equation}\label{eq:lightconeLag}
\begin{aligned}
\mathcal{L}^{(2,\nu)}=&-\frac{1}{2}\Big(-\partial_+\rho\partial_-\sigma-\partial_-\rho\partial_+\sigma-2\chi_{1+}\partial_-\rho-2\chi_{1-}\partial_+\rho-2\chi_{2+}\partial_-\sigma-2\chi_{2-}\partial_+\sigma
\\
&-2\chi_{1+}\chi_{2-}-2\chi_{1-}\chi_{2+}+\rho\text{Tr}[P_+P_-+2P_+v_-+2P_-v_++2v_+v_- ]\Big)+
\mathcal{E}(\nu),
\end{aligned}
\end{equation}
where we will assume from now on that
\begin{align}
\label{eq:sqrtnu} \nu=&\sqrt{\Big(\chi_{1+}\chi_{2+}-\frac{\rho}{2}\text{tr}(v_+^2)\Big)\Big(\chi_{1-}\chi_{2-}-\frac{\rho}{2}\text{tr}(v_-^2)\Big)} \equiv \sqrt{\nu_+\nu_-}\, ,
\\
p=&\chi_{1+}\chi_{2-}+\chi_{1-}\chi_{2+}-\rho\,\text{Tr}(v_+^2)\, .
\end{align}
It was shown in \cite{Baglioni:2025tsc} that the square-rooted version of the variable $\nu_{2}$ from the previous section, defined in \eqref{eq:nu}, is required in order to ensure that the Legendre transform procedure is well-defined, as to be seen momentarily. The auxiliary field algebraic equations stemming from the Lagrangian \eqref{eq:lightconeLag} are
\begin{align}
P_\pm \doteq& -v_\pm-\frac{1}{2}\mathcal{E}_\nu\sqrt{\frac{\nu_\pm}{\nu_\mp}}v_\mp\, ,\\
\partial_\pm\rho\doteq&-\chi_{2\pm}-\frac{1}{2}\mathcal{E}_\nu\sqrt{\frac{\nu_\pm}{\nu_\mp}}\chi_{2\mp}\, ,\\
\partial_\pm\sigma\doteq&-\chi_{1\pm}-\frac{1}{2}\mathcal{E}_\nu\sqrt{\frac{\nu_\pm}{\nu_\mp}}\chi_{1\mp}\,.
\end{align}

We can impose these algebraic equations on the Lagrangian in order to obtain a partially on-shell expression. In particular, after noting that 
\begin{align}
&\nonumber-\tfrac{1}{2}\Big(-\partial_+\rho\partial_-\sigma-\partial_-\rho\partial_+\sigma-2\chi_{1+}\partial_-\rho-2\chi_{1-}\partial_+\rho
\\
&\hspace{4.85cm}-2\chi_{2+}\partial_-\sigma-2\chi_{2-}\partial_+\sigma-2\chi_{1+}\chi_{2-}-2\chi_{1-}\chi_{2+} \Big) 
\\
&\doteq \!-\tfrac{1}{2}\Big(\chi_{1+}\chi_{2-}\!+\!\chi_{1-}\chi_{2+}\!+\!\mathcal{E}_\nu\Bigl(\sqrt{\frac{\nu_-}{\nu_+}}\chi_{1+}\chi_{2+}\!+\!\sqrt{\frac{\nu_+}{\nu_-}}\chi_{1-}\chi_{2-}\Bigr)\!-\!\tfrac{1}{4}\mathcal{E}_\nu^2\big(\chi_{1+}\chi_{2-}\!+\!\chi_{1-}\chi_{2+}\big)\!\Big) \, ,
\notag 
\end{align}
and
\begin{equation}
\begin{aligned}
&-\tfrac{1}{2}\rho\text{Tr}\left[P_+P_-+2 P_+v_-+2 P_-v_++2 v_+v_-\right]
\\
&\doteq -\tfrac{1}{2}\rho\text{Tr}\left[-v_+v_--\tfrac{1}{2}\mathcal{E}_\nu\Bigl(\sqrt{\frac{\nu_-}{\nu_+}}v_+^2+\sqrt{\frac{\nu_+}{\nu_-}}v_-^2\Bigr)+\tfrac{1}{4}\mathcal{E}_\nu^2v_+v_-\right]\, ,
\end{aligned}
\end{equation}
one can check that the Lagrangian can be written as
\begin{align}\label{eq:partialonshellLnu}
\mathcal{L}^{(2,\nu)}\doteq-\frac{1}{2}p
\Big(1-\frac{1}{4}\mathcal{E}_\nu^2\Big)-\nu \mathcal{E}_\nu+\mathcal{E} \,\, .
\end{align} 
One can therefore operate a Legendre transform,
\begin{equation}
 -\nu \mathcal{E}_\nu+ \mathcal{E}=\mathcal{H}(\mu)\, ,\quad \text{with}\quad  \mu=\mathcal{E}_\nu, \quad \text{and} \quad \nu=-\mathcal{H}_{\mu}\equiv -\frac{\partial \mathcal{H}(\mu)}{\partial\mu}\,,
\end{equation}
and obtain the analogue of the Lagrangian \eqref{eq:partialonshellLnu} in the $\mu$-frame,
\begin{align}
    \mathcal{L}^{(2,\,\mu)}\doteq-\frac{p}{2}\Big(1-\frac{\mu^2}{4}\Big)+\mathcal{H}(\mu) \,\, .
\end{align}
Despite the original Lie-algebraic structure of the auxiliary fields $v$ and the dependence of the auxiliary combination $\nu_2$ on the physical field $\rho$, the Legendre transform leads to a theory where the new auxiliary variable $\mu$ is a single scalar oblivious to both the underlying Lie algebra and the physical field $\rho$. This simplification in the structure of the auxiliary fields represents a feature of the applied transformation and, as already observed in \cite{Baglioni:2025tsc}, will be counter-balanced by a more complicated structure of the Lagrangian.

We can then write
\begin{align}
\partial_+\rho\partial_-\sigma+\partial_-\rho\partial_+\sigma-\rho\text{Tr}(P_+P_-) \equiv& I_{(+-)} \doteq p\Bigl(1+\frac{\mu^2}{4}\Bigr)-2\mu \mathcal{H}_{\mu}\, ,
\\
\label{eq:Jpm2}
\!\Bigl(\!\partial_+\rho\partial_+\sigma-\frac{\rho}{2}\text{Tr}(P_+^2)\!\Bigr) \Bigl(\!\partial_-\rho\partial_-\sigma-\frac{\rho}{2}\text{Tr}(P_-^2)\!\Big) \equiv& I_{(++)}I_{(--)} \doteq \Big(\frac{p}{2}\mu-\mathcal{H}_{\mu}\Bigl(1+\frac{\mu^2}{4}\Bigr)\!\Big)^2\, , 
\end{align}
and invert these relations in order to obtain the expressions 
\begin{align}
p\doteq& \frac{4(\mu^2+4)}{(\mu^2-4)^2} \Bigl( I_{(+-)}\pm\frac{8\mu}{(\mu^2+4)}\sqrt{I_{(++)}I_{(--)}}\Bigr)\, ,
\\
\label{eq:Hmueq}\mathcal{H}_{\mu} \doteq& \frac{8\mu}{(\mu^2-4)^2}\Big( I_{(+-)}\pm\frac{4+\mu^2}{2\mu}\sqrt{I_{(++)}I_{(--)}}\Big).
\end{align}
As we are taking the square root of \eqref{eq:Jpm2}, the above formulae come in principle with a leftover arbitrariness in the choice of the + or - sign, which will be fixed later on by demanding consistency. It is then easy to work backwards and reconstruct the appropriate Lagrangian allowing for the recovery of the second equation \eqref{eq:Hmueq},
\begin{equation}
\begin{aligned}\label{eq:muframeLag1}
\mathcal{L}^{(2,\,\mu)}\!=&+\!\Big(\frac{4}{\mu^2-4}\!+\!c_1\Big)\!\Big(\partial_+\rho\partial_-\sigma+\partial_-\rho\partial_+\sigma-\rho\text{Tr}(P_+P_-)\Big)
\\
&+\!\Big(\!\!\mp\frac{4\mu}{\mu^2-4}\!+\!c_2\Big)\sqrt{\!\Big(\partial_+\rho\partial_+\sigma-\frac{\rho}{2}\text{Tr}(P_+^2)\Big)\! \Big(\partial_-\rho\partial_-\sigma-\frac{\rho}{2}\text{Tr}(P_-^2)\Big)} +\mathcal{H}(\mu) \, ,
\end{aligned}
\end{equation}
with $(c_1, c_2)$ constants. Indeed, using the ansatz
\begin{equation}\label{eq:muframelagansatz}
 \mathcal{L}^{(2,\,\mu)} =F(\mu) I_{(+-)}+G(\mu)\sqrt{I_{(++)}I_{(--)}}+\mathcal{H}(\mu) \,\, ,
\end{equation}
the relevant auxiliary field equation \eqref{eq:Hmueq} can be obtained by varying with respect to $\mu$,
\begin{equation}\label{eq:auxmueq}
 \Big(\partial_{\mu}F(\mu) I_{(+-)}+\partial_{\mu}G(\mu)\sqrt{I_{(++)}I_{(--)}}+\mathcal{H}_{\mu}\Big)\delta\mu\doteq0\, ,   
\end{equation}
and directly comparing the $\mu$-dependent prefactors, obtaining the equations 
\begin{equation}
\partial_{\mu}F= -\frac{8\mu}{(\mu^2-4)^2}\,,
\qquad \qquad
\partial_{\mu}G= \pm\frac{4(4+\mu^2)}{(\mu^2-4)^2}\, .
\end{equation}
In turn, these allow to derive the general form of $F$ and $G$
\begin{equation}\label{F-G-functions}
F(\mu)=\frac{4}{\mu^2-4}+c_1\equiv f(\mu)+c_1\,,
\qquad \qquad G(\mu) =\mp\frac{4\mu}{\mu^2-4}+c_2\equiv g(\mu)+c_2\, .
\end{equation}
Demanding the equivalence of the equations of motion in the $\nu$- and $\mu$- frames can then be exploited, as shown in appendix \ref{app:equivalence-of-EOM}, to fix the integration constants to $c_1=\frac{1}{2}$ and $c_2=0$ in \eqref{F-G-functions} and to select the minus sign for $g(\mu)$. This choice determines the final form of the $\mu$-frame Lagrangian \eqref{eq:muframelagansatz}, which reads
\begin{equation}
\begin{aligned}\label{eq:muframeLagfinal}
\mathcal{L}^{(2,\,\mu)}=&+\frac{1}{2}\Big(\frac{\mu^2+4}{\mu^2-4}\Big)\Big(\partial_+\rho\partial_-\sigma+\partial_-\rho\partial_+\sigma-\rho\text{Tr}(P_+P_-)\Big)
\\
&-\Big(\frac{4\mu}{\mu^2-4}\Big)\sqrt{\Big(\partial_+\rho\partial_+\sigma-\frac{\rho}{2}\text{Tr}(P_+^2)\Big) \Big(\partial_-\rho\partial_-\sigma-\frac{\rho}{2}\text{Tr}(P_-^2)\Big)}+\mathcal{H}(\mu)\, .
\end{aligned}
\end{equation}
Finally, as discussed within the $\nu$-frame summary in subsection \ref{subsec:nuframeBM}, the equations of motion arising from \eqref{eq:muframeLagfinal} must be supplemented with the appropriate Virasoro constraints. In order to switch to the $\mu$-frame formulation, one should also assess how the latter change. In the $\nu$-frame  their expression reads as in \eqref{eq:deformedVirasoro}.  
The constraints are derived from an explicit computation of the Geroch algebra cocycles, a computation which relies exclusively on the pole structure of the spectral parameter coefficients in the Lax connection \cite{Paulot:2004hh,Cesaro:2025msv}. Given that this structure is left unchanged in passing from the $\nu$- to the $\mu$-frame, and that the only modification introduced concerns the expressions of the deformed dual currents, the Virasoro constraints in the latter frame read (on the configuration subspace where the auxiliary field equations hold)
\begin{align}\label{eq:muVir}
 \nonumber \tilde{\mathcal{S}}^{(\mu)}_\pm \doteq& \pm\frac{\rho}{4}\text{Tr}\Bigg(\frac{(P_\pm\pm\mathcal{P}^{(\mu)}_\pm)^2}{\partial_\pm\rho\pm\mathcal{R}^{(\mu)}_\pm}-\frac{(P_\pm\mp\mathcal{P}^{(\mu)}_\pm)^2}{\partial_\pm\rho\mp\mathcal{R}^{(\mu)}_\pm}\Bigg) \\
 \nonumber=&\pm\frac{\rho}{4}\text{Tr}\left(\frac{\Big((1\mp1\mp2f(\mu))P_\pm\pm\mu f(\mu)\sqrt{\frac{I_{(\pm\pm)}}{I_{(\mp\mp)}}}P_\mp\Big)^2}{\Big((1\mp1\mp2f(\mu))\partial_\pm\rho\pm\mu f(\mu)\sqrt{\frac{I_{(\pm\pm)}}{I_{(\mp\mp)}}}\partial_\mp\rho\Big)}\right.\\
 &\hspace{4.5cm}\left.-\frac{\Big((1\pm1\pm2f(\mu))P_\pm\mp\mu f(\mu)\sqrt{\frac{I_{(\pm\pm)}}{I_{(\mp\mp)}}}P_\mp\Big)^2}{\Big((1\pm1\pm2f(\mu))\partial_\pm\rho\mp\mu f(\mu)\sqrt{\frac{I_{(\pm\pm)}}{I_{(\mp\mp)}}}\partial_\mp\rho\Big)}\right)\, .
\end{align}
Where now
\begin{equation}
 \tilde{\mathcal{S}}^{(\mu)}_\pm\doteq -\partial_\pm\tilde{\sigma}+2f(\mu)\partial_\pm \sigma +\mu f(\mu)\sqrt{\frac{I_{(\pm\pm)}}{I_{(\mp\mp)}}}\partial_\mp\sigma.
\end{equation}
As a check of the correctness of \eqref{eq:muVir}, one can prove that it leads to \eqref{eq:fieldeq3}, if imposed together with \eqref{eq:fieldeq1} and \eqref{eq:fieldeq2}. Based on the proof in appendix C of \cite{Cesaro:2025msv}, there is no need to perform explicitly the computation: the equations are the very same ones, with the auxiliary fields replaced on shell by \cref{eq:auxofEnurho,eq:auxofEnusigma,eq:auxofEnuP}.
\section{Uplifts to $D=4$}\label{sec:uplifts}
It is a result due to Hermann Weyl \cite{Weyl:1929fm} that the Einstein-Hilbert action can be written in terms of the anholonomicity coefficients,
\begin{equation}
    \Omega_{AB}{}^C=E_A{}^M E_B{}^N(\partial_M E _N{}^C-\partial_N E_M{}^C)\,,
\end{equation}where $E_M{}^A$ is the vierbein, as 
\begin{equation}\label{eq:action0}
    S_{\text{EH}}^{(4)}=\int d^4x \mathcal{L}^{(4)}_{\text{EH}}=-\frac{1}{8}\int d^4 x E\left(\Omega^{ABC}\Omega_{ABC}+2\Omega^{ABC}\Omega_{ACB}-4\Omega_{AB}{}^B\Omega^{AC}{}_C\right)\, ,
\end{equation}
up to boundary contributions.  Written in this form,  \eqref{eq:action0} constitutes the basic formulation of the ``teleparallel-equivalent of general relativity" (TEGR) \cite{Maluf:2013gaa} -- in the \textit{pure tetrad form}, which only constitutes a subsector of the full teleparallel theory -- and it lends itself very efficiently to the construction of a parent action encompassing the dual graviton degrees of freedom \cite{Boulanger:2008nd}. In this work, we will take \eqref{eq:action0} as a natural starting point for the construction of an ansatz for the uplift to $D=4$ of the two-dimensional models \eqref{eq:Ldef} and \eqref{eq:muframeLagfinal}.  One indication that this ansatz is a promising candidate language for describing the uplift of the deformed BM model, is that it seems to automatically encode the invariance under electro-magnetic-like duality transformations of the graviton. This suggests that \eqref{eq:action0} might be amenable to the auxiliary field approach of Ivanov and Zupnik in the construction of deformed electro-magnetic duality-invariant Lagrangians \cite{Ivanov:2001ec ,Ivanov:2002ab,Ivanov:2003uj}. Some hints in this direction are provided in appendix \ref{sec:linearisation}, where we  linearise the action \eqref{eq:action0} and relate it to the linearised action \cite{Henneaux:2004jw,Bunster:2012km}, known to be self-dual. 
\subsection{Uplift of the undeformed model}
As a preliminary check, let us prove that, upon the frame ansatz \eqref{eq:vierbein1}, we recover the correct $D=2$ SL(2) sigma-model coupled to a dilaton and a conformal factor, \eqref{eq:LagMM}. The only non-null components of the anholonomicity coefficients are
\begin{align}
    \Omega_{\alpha\beta}{}^\gamma&=2 e_\alpha{}^\mu e_\beta{}^\nu \partial_{[\mu}e_{\nu]}{}^\gamma\, ,\\
    \Omega_{\alpha b}{}^c&=-\Omega_{b\alpha}{}^c=e_\alpha{}^\mu (\tfrac{1}{2}\delta_{b}{}^{c}\rho^{-1}\partial_\mu\rho+\mathcal{J}_{\mu\, b}{}^c)\, ,
\end{align}
where we have defined the Maurer-Cartan current
\begin{equation}\label{eq:MC-current}
    \mathcal{J}_{\mu\, a}{}^c=(\mathcal{V}^{-1})_a{}^m\partial_\alpha  \mathcal{V}_m{}^c\, ,
\end{equation}
decomposable in the compact and non-compact parts
\begin{equation}
    \mathcal{V}^{-1}\partial_\mu  \mathcal{V}=P_\mu +Q_\mu\,, \quad P_\mu\in\mathfrak{sl}_2\,, \quad Q_\mu\in\mathfrak{so}_2\,.
\end{equation}
Notice that, as an $\mathfrak{sl}_2$ element in the fundamental representation, $\mathcal{J}_{\mu\, a}{}^c$ is traceless in $(a,c)$.

In the conformal gauge $e_\mu{}^\alpha=e^\sigma \delta_\mu{}^\alpha$, the anholonomicity coefficients simplify to
\begin{align}
    \Omega_{\alpha\beta}{}^\gamma&=2 e^{-\sigma} \partial_{[\alpha}\sigma \delta_{\beta]}{}^\gamma\, ,\\
    \Omega_{\alpha b}{}^c&=-\Omega_{b\alpha}{}^c=e^{-\sigma} (\tfrac{1}{2}\delta_{b}{}^{c}\rho^{-1}\partial_\alpha\rho+\mathcal{J}_{\alpha\, b}{}^c)\, ,
\end{align}
with no further distinction being made between curved and flat $D=2$ spacetime indices.

The effective action \eqref{eq:action0} can then be unfolded as
\begin{align}\label{eq:action1}
S^{(2)}_{\text{EH}}=- \frac{1}{8}V_{T^2}\int d^2 x\, \rho e^{2\sigma}\Big(&+\Omega_{\alpha\beta\gamma}\Omega^{\alpha\beta\gamma}+2\Omega_{\alpha\beta\gamma}\Omega^{\alpha\gamma\beta}-4\Omega_{\alpha\gamma}{}^\gamma\Omega^{\alpha\sigma}{}_\sigma\\
&+2\Omega^{\alpha bc}\Omega_{\alpha bc}+2\Omega^{\alpha bc}\Omega_{\alpha cb}-4\Omega_{\alpha d}{}^d\Omega^{\alpha f}{}_f-8\Omega_{\alpha\gamma}{}^\gamma \Omega^{\alpha c}{}_c\Big)\, ,
 \notag
\end{align}
where we have assumed the $T^2$ volume $\int d^2y=V_{T^2}$ (with $y$ the internal coordinates) to be constant (and to be thus effectively set to 1).

A quick computation shows that
\begin{align}
\nonumber \Omega_{\alpha\beta\gamma}\Omega^{\alpha\beta\gamma}&=2e^{-2\sigma}\partial_\alpha \sigma \partial^\alpha \sigma\, ,\\
\nonumber2\Omega_{\alpha\beta\gamma}\Omega^{\alpha\gamma\beta}&= 2e^{-2\sigma}\partial_\alpha \sigma \partial^\alpha \sigma\, ,\\
\nonumber -4\Omega_{\alpha\gamma}{}^\gamma\Omega^{\alpha\lambda}{}_\lambda&=-4e^{-2\sigma}\partial_\alpha \sigma \partial^\alpha \sigma\, ,\\
\nonumber 2\Omega^{\alpha bc}\Omega_{\alpha bc}&=e^{-2\sigma}(\rho^{-2}\partial_\mu\rho\partial^\mu\rho+2\mathcal{J}_\mu{}^{bc}\mathcal{J}^\mu{}_{bc})\, ,\\
\nonumber 2\Omega^{\alpha bc}\Omega_{\alpha cb}&=e^{-2\sigma}(\rho^{-2}\partial_\mu\rho\partial^\mu\rho+2\mathcal{J}_\mu{}^{bc}\mathcal{J}^\mu{}_{cb})\, ,\\
\nonumber -4\Omega_{\alpha d}{}^d\Omega^{\alpha f}{}_f&=-4e^{-2\sigma}\rho^{-2}\partial_\mu\rho\partial^\mu\rho \, ,\\
\nonumber
-8\Omega_{\alpha\gamma}{}^\gamma \Omega^{\alpha c}{}_c&=
-8 e^{-2\sigma}\rho^{-1}\partial_\mu\rho\partial^\mu\sigma\, .
\end{align}
Substituting back into the action \eqref{eq:action1}, such expressions return
\begin{align}\label{eq:undeformedaction}
 \nonumber S^{(2)}_{\text{EH}}&=- \frac{1}{8}\int d^2x\Big(-2\rho^{-1}\partial_\mu\rho\partial^\mu\rho -8\partial_\mu\rho\partial^\mu\sigma+2\rho(\mathcal{J}_\mu{}^{bc}\mathcal{J}^\mu{}_{bc}+\mathcal{J}_\mu{}^{bc}\mathcal{J}^\mu{}_{cb})\Big) \\
 &=\int d^2x\Big( \partial^\mu\rho\partial_\mu\sigma-\frac{\rho}{2}(P_{\mu\,a}{}^bP^\mu{}_b{}^a)\Big)\, .
\end{align}
In the last passage, we have rescaled $\sigma \rightarrow \sigma+\frac{1}{4}\text{ln}\rho$ and further used the fact that
\begin{equation}
  P_{\mu\,a}{}^b=\frac{1}{2}(\mathcal{J}_{\mu\,a}{}^b+\mathcal{J}_{\mu}{}^b{}_a)\, .  
\end{equation}

\subsubsection{Auxiliary field formulation}
Let us introduce an auxiliary field sharing the same algebraic structure as the anholonomicity coefficients,
\begin{equation}
  Y_{AB}{}^C=- Y_{BA}{}^C \,.
\end{equation}
Differently from \cite{Boulanger:2008nd}, we will rewrite the Einstein-Hilbert action in a way which is reminiscent of the auxiliary field deformations for the principal chiral model \cite{Ferko:2024ali}\footnote{There is some arbitrariness in the coefficients $\{4,2\}$ in front of the second and third line. They can be replaced by any $k_1, k_2$ satisfying $1-\frac{k_1^2}{2k_2}+\frac{k_1^2}{4k_2}=-1$.}, 
\begin{align}\label{eq:SEHaux}
 \nonumber   S^{(4)}_{\text{EH}}[\Omega, Y]=+\frac{1}{8}\int d^4 x E\Big(&+\Omega^{ABC}\Omega_{ABC}+2\Omega^{ABC}\Omega_{ACB}-4\Omega_{AB}{}^B\Omega^{AC}{}_C\\
\nonumber&+4(Y^{ABC}\Omega_{ABC}+2Y^{ABC}\Omega_{ACB}-4Y_{AB}{}^B\Omega^{AC}{}_C) \\
&+2(Y^{ABC}Y_{ABC}+2Y^{ABC}Y_{ACB}-4Y_{AB}{}^BY^{AC}{}_C)\Big).
\end{align}
Notice the reversed sign of the first line in comparison to \eqref{eq:action0}. The auxiliary field can be integrated out, returning the algebraic equation of motion
\begin{equation}
  Y_{ABC}= -\Omega_{ABC}\, , 
\end{equation}
which in turn brings us back to \eqref{eq:action0}.

Consider now the same reduction ansatz for the $D=4$ auxiliary field as for the physical anholonomicity coefficients: the non-vanishing components are assumed to be
\begin{equation}
 Y_{ABC}=(Y_{\alpha bc}, Y_{b\alpha c}, Y_{\alpha\beta\gamma})\, ,   
\end{equation}
and an explicit parametrisation of these components in terms of effective $D=2$ auxiliary fields will be given in a moment.
The second line of \eqref{eq:SEHaux} can be explicitly written as
\begin{align}\label{eq:line2}
\nonumber&4E(Y^{ABC}\Omega_{ABC}+2Y^{ABC}\Omega_{ACB}-4Y_{AB}{}^B\Omega^{AC}{}_C)\\
\nonumber&=4\rho e^{2\sigma}\Big(Y^{\alpha bc}\Omega_{\alpha bc}+Y^{b \alpha c}\Omega_{b \alpha c}+Y^{\alpha\beta\gamma}\Omega_{\alpha\beta\gamma}+2Y^{\alpha b c}\Omega_{\alpha cb}+2Y^{\alpha\beta\gamma}\Omega_{\alpha\gamma\beta}\\
\nonumber&\hspace{3.35cm}-4\Omega_{\alpha c}{}^cY^{\alpha d}{}_d-4\Omega_{\alpha c}{}^c Y^{\alpha \gamma}{}_\gamma-4\Omega_{\alpha\gamma}{}^\gamma Y^{\alpha d}{}_d-4\Omega_{\alpha\gamma}{}^\gamma Y^{\alpha\rho}{}_\rho\Big)\\
\nonumber&=4\rho e^{\sigma}\Big(-4Y^{\alpha c}{}_c\partial_\alpha \sigma -(2Y^{\alpha c}{}_c+4Y^{\alpha\gamma}{}_\gamma)\rho^{-1}\partial_\alpha\rho+2Y^{\alpha bc}(\mathcal{J}_{\alpha bc}+\mathcal{J}_{\alpha cb})\Big)\\
&=4\rho e^{\sigma}\Big(-4Y^{\alpha c}{}_c\partial_\alpha \sigma -(2Y^{\alpha c}{}_c+4Y^{\alpha\gamma}{}_\gamma)\rho^{-1}\partial_\alpha\rho+4Y^{\alpha bc}P_{\alpha bc}\Big)\, ,
\end{align}
whereas the last one reads
\begin{align}\label{eq:line3}
\nonumber&2\rho e^{2\sigma}\Big(Y_{\alpha\beta\gamma}Y^{\alpha\beta\gamma}+2Y_{\alpha\beta\gamma}Y^{\alpha\gamma\beta}-4Y_{\alpha\gamma}{}^\gamma Y^{\alpha\sigma}{}_\sigma\\
 &\hspace{2.91cm}+2Y^{\alpha bc}Y_{\alpha bc}+2Y^{\alpha bc}Y_{\alpha cb}-4Y_{\alpha d}{}^dY^{\alpha f}{}_f-8Y_{\alpha\gamma}{}^\gamma Y^{\alpha c}{}_c\Big)\, .
\end{align}
We assume now a KK-reduction ansatz for the auxiliary fields of the following form,
\begin{equation}\label{eq:auxKKansatz}
    Y_{\alpha\beta}{}^{\gamma}=2e^{-\sigma}\tilde{\chi}_{1\;[\alpha}\delta_{\beta]}{}^\gamma\, ,\qquad Y_{\alpha b}{}^c=-Y_{b \alpha}{}^c=e^{-\sigma}\left(\frac{\rho^{-1}}{2}\chi_{2\;\alpha}\delta_b{}^c+\tilde{v}_{\alpha\,b}{}^c\right),
\end{equation}
with $\tilde{v}_{\alpha\,b}{}^c$ having the same algebraic structure as the Maurer-Cartan current $\mathcal{J}_{\mu a}{}^b$ defined in \eqref{eq:MC-current}. Then \eqref{eq:line3} can be shown to take the form
\begin{align}
  8\rho\,v_{\mu\, ab} v^{\mu \,ab}-4\rho^{-1}\chi_{2\;\mu}\chi_2^\mu-16\,\tilde{\chi}_{1\;\mu}\chi_2^\mu \, , 
\end{align}
while \eqref{eq:line2} becomes
\begin{align}
  16 \,\rho P_{\mu\, ab}v^{\mu\, ab}-8\rho^{-1}\partial_\mu\rho\chi_2^\mu-16\,\partial_\mu\rho\tilde{\chi}_1^\mu-16\,\partial_\mu\sigma\chi_2^\mu\, ,
\end{align}
where we have defined
\begin{equation}
    v_{\mu}{}^{ab}=\frac{1}{2}(\tilde{v}_{\mu}{}^{ab}+\tilde{v}_{\mu}{}^{ba})\, .
\end{equation}
Rescaling by an overall factor of $V_{T^2}$ (or, equivalently setting it to 1), the effective Einstein-Hilbert action in $D=2$ \eqref{eq:SEHaux} can be rewritten as
\begin{align}
 \nonumber S^{(2)}_{\text{EH}}&= \int d^2 x \Big(-\partial^\mu\rho\partial_\mu\sigma-\tfrac{1}{4}\rho^{-1}\partial^\mu\rho\partial_\mu\rho+\rho(\tfrac{1}{2}P_{\mu\,a}{}^bP^\mu{}_b{}^a+2P_{\mu\,ab}v^{\mu\,ab}+v_{\mu\,ab}v^{\mu\,ab})\\
 &\hspace{1.9cm}-2(\partial_\mu\rho\tilde{\chi}_1^\mu+\partial_\mu\sigma\chi_2^\mu+\tilde{\chi}_1^\mu\chi_{2\,\mu})-\tfrac{1}{2}\rho^{-1}\chi_{2\,\mu}\chi_2^\mu-\rho^{-1}\partial_\mu\rho\chi_2^\mu\,\Big) .
 \end{align}
By rescaling $\sigma\rightarrow \sigma+\frac{1}{4}\text{ln}\rho$ this further becomes
\begin{align}
 \nonumber S^{(2)}_{\text{EH}}&= \int d^2 x \Big(-\partial^\mu\rho\partial_\mu\sigma+\rho(\tfrac{1}{2}P_{\mu\,a}{}^bP^\mu{}_b{}^a+2P_{\mu\,ab}v^{\mu\,ab}+v_{\mu\,ab}v^{\mu\,ab})\\
 &\hspace{1.9cm}-2(\partial_\mu\rho\tilde{\chi}_1^\mu+\partial_\mu\sigma\chi_2^\mu+\tilde{\chi}_1^\mu\chi_{2\,\mu})-\tfrac{1}{2}\rho^{-1}\chi_{2\,\mu}\chi_2^\mu-\tfrac{1}{2}\rho^{-1}\partial_\mu\rho\chi_2^\mu\,\Big) ,
 \end{align}
 and after an analogue redefinition of $\tilde{\chi}_1^\mu$,
 \begin{equation}\label{chi-1-redefinition}
     \tilde{\chi}_1^\mu=\chi_1^\mu-\tfrac{1}{4}\rho^{-1}\chi_2^\mu\, ,
 \end{equation}
we finally arrive to
\begin{align}\label{eq:rescaledchi}
  \nonumber S^{(2)}_{\text{EH}}&= \int d^2 x \Big(-\partial^\mu\rho\partial_\mu\sigma-2(\partial_\mu\rho\chi_1^\mu+\partial_\mu\sigma\chi_2^\mu+\chi_1^\mu\chi_{2\,\mu})\\
 &\hspace{2cm}+\rho(\tfrac{1}{2}P_{\mu\,a}{}^bP^\mu{}_b{}^a+2P_{\mu\,ab}v^{\mu\,ab}+v_{\mu\,ab}v^{\mu\,ab})\Big),   
\end{align}
which is the auxiliary field action of \cite{Cesaro:2025msv} in the presence of a trivial interaction function.

\subsection{Uplift of deformed models} 

In the previous subsection, we have considered a convenient rephrasing of the four-dimensional Einstein-Hilbert action \eqref{eq:action0}, showing how this can be reduced to the undeformed $D=2$ BM model \eqref{eq:undeformedaction}. This rewriting has further allowed us to show that a $\nu$-frame-like auxiliary field description \eqref{eq:SEHaux} of such a $D=4$ action correctly reduces -- in the absence of an interaction function -- to the $\nu$-frame auxiliary field version  of the $D=2$ BM model \eqref{eq:rescaledchi}. These results represent a first crucial step towards a four-dimensional uplift of both the $\nu$-frame BM model \eqref{eq:lightconeLag} and its $\mu$-frame version \eqref{eq:muframeLagfinal}, which at this point respectively lack a $D=4$ description of the deforming variable $\nu_{2}$ in \eqref{eq:nu} and of the terms under square root in \eqref{eq:muframeLag1}. While such contributions are respectively expressed in terms of $\nu$-frame auxiliary fields and physical fields, the search for their $D=4$ origin  translates into the challenge of uplifting an identical tensorial structure, which will be our aim in the rest of this subsection,

Consider as an instructive example a generic $D=2$ seed theory $\mathring{\mathcal{L}}$, and as its easiest possible deformation in the $\nu$-frame, select $\mathcal{E}(\nu_2)=\nu_2$. At linear order in a perturbative expansion, the action of the auxiliary fields on the $D=2$ seed Lagrangian $\mathring{\mathcal{L}}$ generates deformations by a specific quadratic combination,
\begin{equation}\label{eq:nu2lin}
   \nu_{2\;\text{lin}}=(2\eta^{\mu\rho}\eta^{\nu\sigma}-\eta^{\mu\nu}\eta^{\rho\sigma}) T_{\mu\nu} T_{\rho\sigma} ,
\end{equation}
of the energy-momentum tensor \cite{Ferko:2024ali}, canonically defined (for a flat background metric) in $D=2$ as
\begin{equation}
    T_{\mu\nu}=\frac{\delta \mathring{\mathcal{L}}}{\delta g^{\mu\nu}}-\frac{1}{2}g_{\mu\nu}\mathring{\mathcal{L}}.
\end{equation}

This example suggests to check whether, at linear order in a perturbative expansion, the naive $D=4$ ansatz
\begin{equation}
 \nu_{4\;\text{lin}} =    (2\eta^{AC}\eta^{BD}-\eta^{AB}\eta^{CD}) T_{AB} T_{CD}, 
\end{equation}
with
\begin{equation}\label{eq:ansatzforT}
 T_{A B}=\frac{\delta \mathcal{L}^{(4)}_{\text{EH}}}{\delta \eta^{
 A B}}-\frac{1}{2}\eta_{A B}\mathcal{L}^{(4)}_{\text{EH}},    
\end{equation}
reduces to \eqref{eq:nu2lin} in $D=2$. Of course, $T_{AB}$ will not generically be traceless in $D=4$, as this property relies on the relative coefficient between the first and the second term in \eqref{eq:ansatzforT}, but this should not pose a problem. Building on this intuition one finds that, for $\mathcal{L}^{(4)}_{EH}$ defined as in \eqref{eq:action0}, the tensor $T_{AB}$ in \eqref{eq:ansatzforT} takes the form
\begin{align}\label{eq:defTnot quiet}
  \nonumber T_{AB}=&-\frac{1}{8}E\Big(2\Omega_{ACD}\Omega_{B}{}^{CD}+2\Omega_{ACD}\Omega_B{}^{DC}-4\Omega_{AC}{}^{C}\Omega_{BD}{}^{D}-\Omega_{CDA}\Omega^{CD}{}_{B}\Big)\\
  &-\frac{1}{2}\eta_{AB}\mathcal{L}^{(4)}_{\text{EH}},
\end{align}
which is manifestly symmetric in the lower indices. The different blocks of the associated KK-reduction are given as follows: the first one reads
\begin{align}
2\Omega_{ACD}\Omega_B{}^{CD}=2(\Omega_{\alpha CD}\Omega_\beta{}^{CD}, \Omega_{a CD}\Omega_b{}^{CD})\, ,   
\end{align}
with the $\alpha\beta$ and $ab$ components
\begin{align}
 2\Omega_{\alpha CD}\Omega_\beta{}^{CD} =&  2e^{-2\sigma}\Big(\eta_{\alpha\beta}\partial_\gamma\sigma\partial^\gamma\sigma+\frac{1}{2}\partial_\alpha\text{ln}\rho\partial_\beta\text{ln}\rho+\mathcal{J}_{\alpha bc}\mathcal{J}_\beta{}^{bc}\Big)\, ,\\
 2\Omega_{a CD}\Omega_b{}^{CD}=&2e^{-2\sigma}\Big(\frac{1}{4}\eta_{ab}\partial_\gamma\text{ln}\rho\partial^\gamma\text{ln}\rho+P_{\gamma ab}\partial^\gamma\text{ln}\rho+\mathcal{J}_{\alpha (a|d}\mathcal{J}^\alpha{}_{|b)}{}^d\Big)\, .
\end{align}
The second and third blocks read
\begin{align}
  2\Omega_{ACD}\Omega_B{}^{DC}=&(2\Omega_{\alpha CD}\Omega_\beta{}^{DC},0)
  \notag \\
  =&\Bigg(2 e^{-2\sigma}\Big(\partial_\alpha\sigma\partial_\beta\sigma+\frac{1}{2}\partial_\alpha\text{ln}\rho\partial_\beta\text{ln}\rho+\mathcal{J}_{\alpha bc}\mathcal{J}_\beta{}^{cb}\Big),0\Bigg)\,,\\
\nonumber-4\Omega_{AC}{}^C\Omega_{BD}{}^D=&\Big(-4\Omega_{\alpha C}{}^C\Omega_{\beta D}{}^D,0\Big) \\=&\Bigg(-4e^{-2\sigma}\Big(\partial_\alpha\sigma\partial_\beta\sigma+\partial_\alpha\text{ln}\rho\partial_\beta\text{ln}\rho+2\partial_{(\alpha}\sigma\partial_{\beta)}\text{ln}\rho\Big),0\Bigg),
\end{align}
while the fourth one 
\begin{equation}
-\Omega_{CDA}\Omega^{CD}{}_{B} =-\Big(\Omega_{CD\alpha}\Omega^{CD}{}_\beta, \Omega_{CDa}\Omega^{CD}{}_b\Big)\, ,
\end{equation}
exhibits  $\alpha\beta$  and $ab$ components
\begin{align}
 -\Omega_{CD\alpha}\Omega^{CD}{}_\beta=&-2 e^{-2\sigma}\Big(\eta_{\alpha\beta}\partial_\gamma\sigma\partial^\gamma\sigma-\partial_\alpha\sigma\partial_\beta\sigma\Big) \, ,\\
 -\Omega_{CD a}\Omega^{CD}{}_b=&-2e^{-2\sigma}\Big(\frac{1}{4}\eta_{ab}\partial_\gamma\text{ln}\rho\partial^\gamma\text{ln}\rho+P_{\gamma ab}\partial^\gamma\text{ln}\rho+\mathcal{J}_{\alpha d(a}\mathcal{J}^\alpha{}^d{}_{b)}\Big)\, \, .
\end{align}

Thus, combining all the terms, and rescaling $\sigma$ as $\sigma \rightarrow \sigma +\frac{1}{4}\text{ln}\rho$, we obtain 
\begin{equation}
  T_{AB}=(T_{\alpha\beta}, T_{ab}) \, , 
\end{equation}
with
\begin{align}\label{Talphabeta-traceless}
T_{\alpha\beta}=-\frac{1}{2}\rho\Big(P_{\alpha bc}P_\beta{}^{bc}-2\partial_{(\alpha}\sigma\partial_{\beta)}\text{ln}\rho\Big)-\frac{1}{2}\eta_{\alpha\beta}\Big(\partial^\mu\rho\partial_\mu\sigma-\frac{\rho}{2}(P_{\mu\,a}{}^bP^\mu{}_b{}^a)\Big)\, .  
\end{align}
The external component block is traceless as desired, and in lightcone coordinates reads
\begin{align}
 T_{\pm\pm}=&  \Big(\partial_\pm\sigma\partial_\pm\rho-\frac{\rho}{2}\text{Tr}(P_\pm^2)\Big)\ ,
\end{align}
while the off-diagonal components $T_{+-}$ and $T_{-+}$ automatically vanish.

However, the object $ T_{AB}$ features a non-vanishing internal block,
\begin{equation}\label{eq:Tint}
T_{ab}= \frac{1}{2}\rho[P_\alpha,Q^\alpha]_{ab}-\frac{1}{2}\delta_{ab}\Big(\partial^\mu\rho\partial_\mu\sigma-\frac{\rho}{2}(P_{\mu\,c}{}^dP^\mu{}_d{}^c)\Big),   
\end{equation}
which returns further non vanishing contributions from the $D=4$ contraction, of the kind
\begin{equation}
  (2\delta^{ab}\delta^{cd}-\delta^{ac}\delta^{bd})T_{ac}T_{bd},  
\end{equation}
in addition to mixing terms between external and internal components. We remind here that the correct ansatz should instead reduce to \eqref{eq:nu} upon compactifying to $D=2$.

This undesired feature can however be easily fixed. Consider as the fundamental object
\begin{equation}
W_{C A}{}^{B}=E_C{}^QE_{A}{}^M\partial_Q E_{M}{}^{B}=
e^{-\sigma}\begin{pmatrix}
    \delta_\alpha{}^\beta\partial_\mu\sigma & 0\\
    0& \frac{1}{2}\delta_{a}{}^{b}\partial_\mu\text{ln}\rho+\mathcal{J}_{\mu\; a}{}^{b}\, 
\end{pmatrix} ,  
\end{equation}
rather than its antisymmetrised version in the first two indices, $\Omega_{AB}{}^C$.
By defining its symmetric and antisymmetric combinations in the last two indices, 
\begin{equation}
 S_{C AB}\equiv 2W_{C (AB)}\, ,\qquad \qquad A_{C AB}\equiv 2W_{C[AB]}\, ,   
\end{equation}
one can show that the commutator 
\begin{align}\label{eq:additionalcomm}
[S_C, A^C]_{AB}=[S_\mu, A^\mu]_{AB}=S_{\mu\; A}{}^CA^\mu{}_{CB}- A_{\mu\; A}{}^CS^\mu{}_{CB}=\begin{pmatrix}
    0&0\\
    0&4 e^{-2\sigma}[P_\mu, Q^\mu]_{ab}\, 
\end{pmatrix},  
\end{align}
has exactly the same structure as the internal commutator in \eqref{eq:Tint}. Therefore, supplementing \eqref{eq:defTnot quiet} with an appropriately weighted contribution \eqref{eq:additionalcomm}
\begin{equation}\label{eq:finalt}
 t_{AB}=T_{AB}-\frac{1}{8}E[S_\mu,A^\mu]_{AB} \,\, ,
\qquad 
\text{with} \qquad
E=\rho e^{2\sigma} \,\, ,
\end{equation}
furnishes the correct uplift. We will now show this is indeed the case. Compute 
\begin{align}
 (2\eta^{AB}\eta^{CD}-\eta^{AC}\eta^{BD})t_{AC}t_{BD}=(2\eta^{\alpha\beta}\eta^{\gamma\sigma}t_{\alpha\gamma}t_{\beta\sigma}+2\delta^{ab}\delta^{cd}t_{ac}t_{bd}-\delta^{ab}\delta^{cd}t_{ab}t_{cd})\, ,   
\end{align}
where we have used the tracelessness of $t_{\alpha\beta}$. The second and the third term on the right hand side, for the specific dimensionality under analysis, $D=2$, nicely compensate each other, and the overall contraction amounts to
\begin{equation}
(2\eta^{AB}\eta^{CD}-\eta^{AC}\eta^{BD})t_{AC}t_{BD}=2\eta^{\alpha\beta}\eta^{\gamma\sigma}t_{\alpha\gamma}t_{\beta\sigma} \,\, .  
\end{equation}
At this point, noting that for any symmetric tensor $
\Xi_{\alpha\beta}$ in $D=2$ and its traceless version $\tilde{\Xi}_{\alpha\beta}=\Xi_{\alpha\beta}-\frac{1}{2}\eta_{\alpha\beta}\eta^{\rho\sigma}\Xi_{\rho\sigma}$, the following identity holds true
\begin{align}\label{tensor-identity}
2\eta^{\alpha\beta}\eta^{\gamma\sigma}\tilde{\Xi}_{\alpha\gamma}\tilde{\Xi}_{\beta\sigma}=   (2\eta^{\alpha\beta}\eta^{\gamma\sigma}-\eta^{\alpha\gamma}\eta^{\beta\sigma})\Xi_{\alpha\gamma}\Xi_{\beta\sigma} \,\, ,
\end{align}
applying \eqref{tensor-identity} to the traceless tensor $t_{\alpha\beta}=T_{\alpha\beta}$ in \eqref{Talphabeta-traceless}, leads us to the desired result, 
\begin{align}\label{eq:nu4linuplift}
\nu_{4\;\text{lin}}= (2\eta^{AB}\eta^{CD}-\eta^{AC}\eta^{BD})t_{AC}t_{BD}= (2\eta^{\alpha\beta}\eta^{\gamma\sigma}-\eta^{\alpha\gamma}\eta^{\beta\sigma})\tau_{\alpha\gamma}\tau_{\beta\sigma}\, ,  
\end{align}
for
\begin{align}
  \tau_{\alpha\beta}=  \rho\Big(\partial_{(\alpha}\sigma\partial_{\beta)}\text{ln}\rho-\frac{1}{2}P_{\alpha bc}P_\beta{}^{bc}\Big).
\end{align}

As mentioned in the beginning of this section, not only does \eqref{eq:nu4linuplift} furnish the uplift of the linearised version of $\nu_2$, but also the uplift of the argument of the square root in \eqref{eq:muframeLagfinal}. The full non-linear version of $\nu_2$ in terms of auxiliaries, on the other hand, is obtained by replacing, using the definitions below, $W_{AB}{}^C$ with $Z_{AB}{}^C$ and $\Omega_{AB}{}^C$ with $Y_{AB}{}^C$ everywhere in \eqref{eq:nu4linuplift}. This allows us to straightforwardly obtain the uplift of the $D=2$ theory in both frames, which represents one of the main results of the paper, and which we hereby display.\\[2pt]

The uplift of the $D=2$ action in the $\nu$-frame
\begin{equation}
 S^{(2,\;\nu)}=\int d^2x   \mathcal{L}^{(2,\;\nu)}\, ,
\end{equation}
with $\mathcal{L}^{(2,\;\nu)}$ defined in
\eqref{eq:Ldef}, is
\begin{align}\label{eq:finalLagnuframe}
  S^{(4,\; \nu)}=S^{(4)}_{\text{EH}}[W, Z]+\int d^4x \;\mathcal{E}(\nu_4).
\end{align}
Here, $S^{(4)}_{\text{EH}}[W, Z]$ is \eqref{eq:SEHaux}, with $\Omega$ and $Y$ expressed in terms of $W$ and $Z$ as
\begin{align}
\Omega_{AB}{}^{C}=2W_{[AB]}{}^C\, ,\qquad \qquad
Y_{AB}{}^{C}&=2Z_{[AB]}{}^C\, ,   
\end{align}
and $\mathcal{E}(\nu_4)$ is the $D=4$ version of $\mathcal{E}(\nu_{2})$ in \eqref{eq:Ldef}. It is a generic function of the variable
\begin{align}\label{eq:nu4}
\nu_4=(2\eta^{AB}\eta^{CD}-\eta^{AC}\eta^{BD})V_{AC}V_{BD}\,  ,  
\end{align}
with $V_{AB}$ symmetric in $(A,B)$ and explicitly written in terms of $D=4$ auxiliary fields
\begin{align}
\nonumber V_{AB}=\frac{E}{8}\Bigg[-&\Big(2Y_{ACD}Y_{B}{}^{CD}+2Y_{ACD}Y_B{}^{DC}-4Y_{AC}{}^{C}Y_{BD}{}^{D}+Y_{CDA}Y^{CD}{}_{B}\Big)\\
 \nonumber +&\frac{1}{2}\eta_{AB}\Big(Y_{CDE}Y^{CDE}+2Y_{CDE}Y^{CED}-4Y_{CD}{}^{D}Y^{CE}{}_{E}\Big)\\
+&2\Big(Z_{CAD}Z^{C}{}_{E}{}^{D}-Z_{CDA}Z^{CD}{}_E\Big)\Bigg].  
\end{align}

Similarly, the uplift of the $D=2$ action in the $\mu$-frame, \eqref{eq:muframeLagfinal}, reads
\begin{align}\label{eq:fullmuframeLag}
&S^{(4,\;\mu)}=\int d^4x \Bigg[\frac{1}{2}\Big(\frac{\mu^2+4}{\mu^2-4}\Big)\mathcal{L}_{(4)}[E]-\Big(\frac{4\mu}{\mu^2-4}\Big)\sqrt{\mathcal{K}[E]}+\mathcal{H}(\mu)\,\Bigg] ,
\end{align}
with 
\begin{equation}
  \mathcal{K}[E]=(2\eta^{AB}\eta^{CD}-\eta^{AC}\eta^{BD})t_{AC}t_{BD} \,\, . 
\end{equation}
$t_{AB}$ is symmetric in $AB$, depends on physical fields and is here reported for convenience,
\begin{align}
\nonumber t_{AB}=\frac{E}{8}\Bigg[-&\Big(2\Omega_{ACD}\Omega_{B}{}^{CD}+2\Omega_{ACD}\Omega_B{}^{DC}-4\Omega_{AC}{}^{C}\Omega_{BD}{}^{D}+\Omega_{CDA}\Omega^{CD}{}_{B}\Big)\\
 \nonumber +&\frac{1}{2}\eta_{AB}\Big(\Omega_{CDE}\Omega^{CDE}+2\Omega_{CDE}\Omega^{CED}-4\Omega_{CD}{}^{D}\Omega^{CE}{}_{E}\Big)\\
+& 2\Big(W_{CAD}W^{C}{}_{B}{}^{D}-W_{CDA}W^{CD}{}_B\Big)\Bigg].    
\end{align}

\subsection{Candidate deforming tensors and residual
symmetries}\label{sec:symmetries}

It is interesting to compare our ansatz \eqref{eq:finalt} to the structure of the energy-momentum tensor for the gravitational field in the teleparallel-equivalent formulation of general relativity (TEGR), canonically defined in \cite{Maluf:2013gaa} and proportional, in the flat basis, to
\begin{align}\label{eq:ttele}
\nonumber &t^{(\text{TEGR})}_{AB}\sim -E\Big(2\Omega_{A}{}^{CD}\Omega_{BCD}+2\Omega_{ACD}\Omega_{B}{}^{DC}-4\Omega_{AC}{}^C\Omega_{BD}{}^D\\
 &\hspace{5.05cm}-4\Omega_{CD}{}^D\Omega^C{}_{BA}-2\Omega^{CD}{}_A\Omega_{BCD}\Big) -\frac{1}{2}\eta_{AB}\mathcal{L}\, .
\end{align}

Given that in $D=2$ the coupling to the auxiliary fields is built to induce deformations by functions of the energy-momentum tensor, it is reasonable to expect that the uplifts to $D=4$ of the deformed BM model should correspond to deformations of general relativity by some four-dimensional appropriate notion of such a tensor. In the teleparallel language, which has shown amenability to the introduction of auxiliary fields and exhibits by construction the  candidate \eqref{eq:ttele}, it thus seems natural to expect a $D=4$ deformation driven by functions of such a tensor. An important thing to notice, however, is that \eqref{eq:ttele} is not a symmetric tensor in its two indices, to begin with. Additionally, while the whole first line and last term on the second line of \eqref{eq:ttele} are contained in our definition \eqref{eq:finalt} of $t_{AB}$, the two tensors exhibit some differences. In particular, their equality would require the first two terms on the second line of \eqref{eq:ttele} to account for the terms $-E\Omega_{CDA}\Omega^{CD}{}_{B}-E[S_\mu,A^\mu]_{AB}$ in \eqref{eq:finalt}, but one can check that this is not the case. Indeed, substituting
\begin{equation}
\begin{aligned}
-4\Omega_{CD}{}^D&\Omega^C{}_{\beta\alpha}-2\Omega^{CD}{}_\alpha\Omega_{\beta CD}
\\
&=2e^{-2\sigma}(\partial_\alpha\sigma\partial_\beta\sigma-\eta_{\alpha\beta}\partial_\gamma\sigma\partial^\gamma\sigma)+4e^{-2\sigma}(\partial_\beta\sigma\partial_\alpha\text{ln}\rho-\eta_{\alpha\beta}\partial_\gamma\sigma\partial^\gamma\text{ln}\rho)\, ,   
\end{aligned}
\end{equation}
in \eqref{eq:finalt} as a replacement for $-E\Omega_{CD\alpha}\Omega^{CD}{}_{\beta}-E[S_\mu,A^\mu]_{\alpha\beta}$, spoils the balance between the coefficients and prevents the $\partial_\alpha\text{ln}\rho\partial_\beta\text{ln}\rho$ contribution to disappear after the necessary rescaling $\sigma\rightarrow \sigma +\tfrac{1}{4}\text{ln}\rho$. Even by symmetrising  \eqref{eq:ttele} in $\alpha \beta$, this behaviour is not improved. We have checked in appendix \ref{sec:appendixgeometry} whether \eqref{eq:finalt} corresponds to another potential candidate for the notion of energy-momentum (pseudo-)tensor in general relativity, namely the Landau-Lifshitz pseudo-tensor \cite{Landau:1975pou}, finding once again a negative answer.

Beyond the difficulties in relating \eqref{eq:finalt} to an appropriate notion of known $D=4$ energy-momentum (pseudo)tensor, a baffling feature of the models \eqref{eq:finalLagnuframe} and \eqref{eq:fullmuframeLag}, is their apparent lack of manifest symmetries. The models do not seem invariant, at least at face-value, under either local Lorentz or general coordinate transformations. This is ultimately due to the behaviour of the quantities $\nu_4$ and $\mathcal{K}$ under both transformations (for the case of diffeomorphisms, this behaviour is computed in appendix \ref{sec:appendixsymmetries}). This in turn implies that the lack of manifest symmetries affects the $\mu$-frame model at an even more fundamental level than just its deformation. 
Indeed, considering the Lagrangian in the $\nu$-frame, for $\mathcal{E}(\nu_4)=0$, on the one hand, we get back the Einstein-Hilbert Lagrangian, 
\begin{equation}
  S^{(4,\; \nu)}\Big\lvert_{\mathcal{E}(\nu_4)=0}=S^{(4)}_{\text{EH}}[W, Z],  
\end{equation}
which is perfectly diffeomorphic-invariant, even before integrating out the auxiliary fields $Z_{AB}{}^C$. On the other hand, by looking at the $\mu$-frame Lagrangian under similar conditions, namely setting $\mathcal{H}(\mu)=0$,
\begin{equation}
  S^{(4,\;\mu)}\Big\lvert_{\mathcal{H}(\mu)=0}=\int d^4 x\Bigg[\frac{1}{2}\Big(\frac{\mu^2+4}{\mu^2-4}\Big)\mathcal{L}_{(4)}[E]-\Big(\frac{4\mu}{\mu^2-4}\Big)\sqrt{\mathcal{K}}\Bigg],   
\end{equation}
after integrating out the auxiliary field $\mu$, we obtain
\begin{equation}\label{eq:H0action}
    S^{(4,\;\mu)}\Big\lvert_{\mathcal{H}(\mu)=0}\doteq\int d^4x\Bigg[\frac{1}{2}\sqrt{\mathcal{L}_{(4)}^2[E]-4\mathcal{K}}\Bigg]\, ,
\end{equation}
which by itself is not diffeomorphic-invariant in full generality. Indeed, based on the computations of appendix \ref{sec:appendixsymmetries}, we find that under a diffeomorphism parametrised by $\xi^M$, the action \eqref{eq:H0action} transforms as
\begin{align}
\nonumber&\delta_\xi S^{(4,\;\mu)}\Big\lvert_{\mathcal{H}(\mu)=0} =\int d^4x\partial_P\Bigg(\xi^P\frac{1}{2}\sqrt{\mathcal{L}_{(4)}^2[E]-4\mathcal{K}}\Bigg)\\ 
&
-\int d^4x\frac{2}{\sqrt{-4\mathcal{K}+\mathcal{L}_{(4)}^2[E]}}\Bigg(\mathcal{K}[E]\partial_P\xi^P+4\eta^{AB}\eta^{CD}E\, t_{AC}\,\tilde{\delta}_\xi[S_\mu, A^\mu]_{BD}\Bigg).
\end{align}
This means that already for \textit{trivial} interaction function, the $\mu$-frame Lagrangian is \textit{not} diffeomorphic-invariant anymore, differently from its $\nu$-frame counterpart. We note that the action \eqref{eq:H0action}, in agreement with the observation in \cite{Baglioni:2025tsc}, already corresponds itself to a deformation of the Einstein-Hilbert action \eqref{eq:EH}, which, on the other hand, is recovered by sending $\mu  \rightarrow 0$.

\section{Conclusions and outlook}\label{sec:conclusions}
In this paper we have focused on the integrable auxiliary field deformation of the BM model derived in \cite{Cesaro:2025msv}; we have constructed the alternative $\mu$-frame formulation of the theory, thus advancing the program initiated in \cite{Baglioni:2025tsc} of extending the latter to the known auxiliary field deformations based on the framework of \cite{Ferko:2024ali}, and we have furthermore succeeded in explicitly uplifting the model to $D=4$ in both the $\nu$- and $\mu$-frames. 
Our results raise some pressing questions, together with further interesting research directions, which we proceed to highlight in this section.

An immediately puzzling feature of the uplift is beyond any doubt its lack of manifest local Lorentz and diffeomorphism invariance. As far as local Lorentz invariance is concerned, to some extent this should have been expected: we worked in the so-called \textit{pure tetrad} formalism, a particular sub-class of the full teleparallel approach in which the ``Weitzenb\"ock" gauge is chosen (whereby the spin connection is set to zero). As remarked in \cite{Krssak:2018ywd}, demanding Lorentz invariance would hence amount to demanding explicit gauge invariance of a gauge theory \textit{after} having committed to a specific gauge, which is of course an unreasonable request. Thus, we expect the theory to become manifestly invariant  under such a symmetry, once expressed through a fully covariant approach which departs from the Weitzenb\"ock gauge and explicitly reintroduces the spin connection (along the lines of \cite{Krssak:2018ywd}). 

The problem of diffeomorphism invariance is a more subtle one. As observed in section \ref{sec:symmetries}, the diffeomorphism group seems broken already in the $\mu$-frame for the $D=4$ model with trivial interaction function. Here as well, with hindsight, the lack of full diffeomorphism invariance is perhaps not such an outrageous outcome. The notion of energy of the gravitational field in general relativity is usually encoded in the form of a pseudo-tensor, which does not transform covariantly (in the literature, this expresses the well known problem of a \textit{local notion of energy \footnote{Or rather,  lack thereof.}} for a gravitational field \cite{Coller:1958tx}). The auxiliary field deformations of \cite{Ferko:2024ali} rely precisely on such a notion of energy-momentum tensor, (which would be only a pseudo-tensor in this case) so it is natural to expect  deformations originated by functions of an object which does not transform covariantly.  

On the other hand, this feature becomes rather perplexing when trying to relate it with the presence of the full affine Geroch symmetry, still present upon dimensional reduction to $D=2$. It seems indeed unrealistic that such an infinite dimensional symmetry in $D=2$ could arise from only a $D=4$ SL(4) (see appendix \ref{sec:appendixsymmetries}), and it is in our opinion a question of the utmost interest to understand its exact higher-dimensional origin in this case, which we hope to tackle in the future.

It should be mentioned that, despite the present work's lack of a satisfactory answer to the question of diffeomorphism invariance, possible further  explanations to this conundrum do exist. In particular, as commented upon in appendix \ref{sec:appendixsymmetries}, the theory could be invariant (order-by-order in a coupling parameter expansion) under a deformed version of the diffeomorphisms, in the spirit of the BRST cohomology approach of \cite{Henneaux:1997bm,Barnich:1995ap}. Furthermore, while we deem unlikely the existence of a different, proper tensor (or tensor density) $\tilde{t}_{AB}$ reducing in $D=2$ to the same components as those of \eqref{eq:finalt}, we cannot rule out instead the existence of a proper tensor (or tensor density) $\tilde{t}_{AB}$ whose total contraction $(2\eta^{AB}\eta^{CD}-\eta^{AC}\eta^{BD})\tilde{t}_{AC}\tilde{t}_{BD}$ reduces precisely to \eqref{eq:nu4linuplift}. In this sense, our result \eqref{eq:finalt} would represent a ``minimal" ansatz and the deformations expressed in terms of such new objects would restore invariance under diffeomorphisms (or transverse diffeomorphisms). We mention that another option to restore, at least formally, diffeomorphism invariance would be the introduction of St\"uckelberg compensating fields, which would allow to regard our theory as a gauge-fixed\footnote{Where the gauge choice breaks diffeomorphisms.} version of a gauge-invariant one. However the resulting physical implications are not clear to us in this case.

Tied to the absence of symmetries, is the presence of ghost modes. Notwithstanding its puzzling features, the teleparallel formulation of the model has the built-in property (shared with the auxiliary field deformation framework) of never introducing field equations with derivatives higher than second order\footnote{In particular, this is one of the properties which makes the modified $f(T)$-theories of gravity \cite{Ferraro:2006jd,Ferraro:2008ey} (where $T$ stands for ``torsion") qualitatively different from the analogue $f(R)$ modifications.}:
this guarantees the theory to be at least free of Ostrogradsky-like ghost instabilities \cite{Ostrogradsky:1850fid,Woodard:2015zca}. We stress again that this peculiar appearance of ``selected" higher-derivatives in $D=4$ could provide a possible explanation to the persistence of the affine symmetry in lower dimensions, contrary to the more customary cases~\cite{Michel:2007vh, Lambert:2006he, Lambert:2006ny, Bao:2007er}, where higher-derivatives in higher dimensions break the duality group. However, even if full local Lorentz invariance was restored through the use of a covariant approach, the theory would still be affected by Boulaware-Deser-like ghosts, which could only be removed via the presence of additional symmetries.

Finally, we conclude this section by highlighting several appealing lines of possible future investigation related to our findings.

Along the $\mu$-frame direction, the expression of the undeformed action \eqref{eq:H0action} is tantalizingly reminiscent of the Born-Infeld formulations of gravitational theories \cite{Deser:1998rj}. Once the absence of diffeomorphism invariance is understood, it would be intriguing to explore further this analogy, and to assess whether its deformations could be related to similar deformations of Born-Infeld gravitational theories in the teleparallel language \cite{Ferraro:2008ey}.

The latter has played in this work a distinguished role in being able to uplift to $D=4$ the integrable deformations \cite{Cesaro:2025msv} of the two-dimensional BM model. In light of this, it also seems natural to investigate possible connections with a recently proposed class of gravitational instantonic solutions \cite{Krssak:2024vzo}, exhibiting strong resemblance to the analogous ones arising in the self-dual sector of Yang-Mills theory and in turn to integrability \cite{Ward:1985gz,Mason:1991rf}.

In the spirit of recent analyses \cite{Bielli:2026ggs}, it would furthermore be appealing to study if and how the existence of a classical Yangian symmetry underlying the BM model \cite{Korotkin:1997fi} would survive the coupling to auxiliary fields, also understanding its relation to the Geroch group and how this might potentially be modified.

Finally, it was recently shown in \cite{Ashwinkumar:2026dwd} how to embed the Yang-Baxter-like deformations of the BM model, also worked out in \cite{Cesaro:2025msv}, within the $D=4$ Chern-Simons framework of \cite{Cole:2023umd}. It would be fascinating to understand whether such an embedding is viable also for the $D=2$ auxiliary deformations here explored, and in particular its relations to the $D=4$ uplifts obtained in this work. This could potentially help in clarifying the observed pathologies that the uplifts seem to be affected by, and could shed further light onto their apparent lack of symmetry.

\section*{Acknowledgements}
We are indebted to Christian Ferko for collaboration on initial stages of this work and for providing useful comments on the draft. It is a pleasure to thank Guillaume Bossard, Axel Kleinschmidt and Henning Samtleben for enlightening discussions on the topic, and in particular Guillaume Bossard for feedback and Axel Kleinschmidt for carefully reviewing the manuscript. M.C. would also like to thank Emiel Claasen and Bastien Duboeuf for helpful discussions, and David Osten both for collaboration on related projects and for pointing at useful literature.
D.B. thanks Nicola Baglioni, Michele Galli, Gabriele Tartaglino-Mazzucchelli for collaboration on topics related to this work.
D.B. has been supported by a Young Scientist Training (YST) Fellowship from Asia Pacific Center for Theoretical Physics (APCTP) and the Thailand NSRF via PMU-B, grant number B13F680083.
We finally acknowledge the ``Workshop on Higher-d Integrability in
Favignana, 2025", where preliminary discussions about the project took place.

%
%
%
%
%
\newpage
\appendix
\section{$\mathbb{Z}_2$ coset models}\label{secion:appendixcoset}
In symmetric spaces $G/K$, the global isometry $G$ has a Lie algebra $\mathfrak{g}$ decomposing as
\begin{align}
\label{eq:kp}
\mathfrak{g}= \mathfrak{g}^{(0)} \oplus \mathfrak{g}^{(1)} = \mathfrak{k} \oplus  \mathfrak{p}\, .
\end{align}
This decomposition is a $\mathbb{Z}_2$-grading of $\mathfrak{g}$ and $\mathfrak{k}$ is the Lie algebra of a (connected) subgroup $K\subset G$, fixed by an involution. 

The coset $\sigma$-model depends on a field $\mathbb{R}^{1,1} \to G/K$, where we refer to $\mathbb{R}^{1,1}$ as (flat) space-time.
We can choose a coset representative $\mathcal{V}(x)\in G$ 
that transforms as
\begin{align}
\label{eq:costrm}
\mathcal{V}(x) \to \mathcal{V}^\prime(x) =g^{-1}\mathcal{V}(x) k(x)\, ,
\end{align}
with global $g\in G$ and local (in space-time) $k(x) \in K$, corresponding to the freedom of choosing a coset representative of $G/K$ at each space-time point. Fixing a specific form of a representative for $G/K$  determines $k(x)$ as a compensating transformation that depends on $g$ and $\mathcal{V}(x)$. In the following and in the main text we have mostly suppressed the dependence on the space-time coordinate $x^\mu$ to avoid cluttering.

The $\mathfrak{g}$-valued Maurer--Cartan current decomposes  according to~\eqref{eq:kp} as
\begin{align}
\label{eq:QP}
\mathcal{V}^{-1}\partial_\mu  \mathcal{V}=P_\mu +Q_\mu\,, \quad P_\mu\in\mathfrak{p}\, ,\; Q_\mu\in\mathfrak{k} .
\end{align}
The two components respectively transform covariantly and as a connection, under~\eqref{eq:costrm}:
\begin{align}
\label{eq:QPtrm}
P_\mu \to  k^{-1}P_\mu k\,, \quad
Q_\mu  \to k^{-1}Q_\mu k+ k^{-1}\partial_\mu k\,.
\end{align}
From the definition~\eqref{eq:QP} one can deduce the Bianchi identities
\begin{align}
\label{eq:Bianchis}
2\partial_{[\mu}Q_{\nu]}+[Q_\mu,Q_\nu]=-[P_\mu,P_\nu]\, , \hspace{15mm} 
D_{[\mu}P_{\nu]}=0\, ,
\end{align}
where we have introduced the $K$-covariant derivative 
\be\label{eq:covder}
D_\mu = \partial_\mu +Q_\mu.
\ee

\section{Equivalence of the EOM in the $\nu$- and $\mu$- frames}\label{app:equivalence-of-EOM}
In this appendix we show the equivalence of the field equations in the two frames, exploiting this to fix the functions $F(\mu)$ and $G(\mu)$ in \eqref{F-G-functions} and in turn the final form of the $\mu$-frame Lagrangian \eqref{eq:muframeLagfinal}. We first look at the equations of motion from the $\nu$-frame perspective, namely considering \cref{eq:Boxrho=smth,eq:dynamicaleom,eq:Boxsigma+smth,eq:algebraicP,eq:algebraicrho,eq:algebraicsigma} expressed in lightcone coordinates and properly rewritten taking into account the definition of $\nu$ in \eqref{eq:sqrtnu}:
\begin{align}
0=&+\partial_+\partial_-\rho+\partial_+\chi_{2\,-}+\partial_-\chi_{2\,+} \, , \label{eq:fieldeq1}
\\
\nonumber 2\rho([v_{-},P_{+}]+[v_{+},P_{-}])=& + \partial_+\rho(P_-+2v_-)+\partial_-\rho(P_++2v_+)
\\ 
\label{eq:fieldeq2} &+ \rho\Big(D_+(P_-+2v_-)+D_-(P_++2v_+)\Big) \, ,
\\
\nonumber \partial_+\partial_-\sigma +\partial_+\chi_{1\,-}+\partial_-\chi_{1\,+}=&-\text{Tr}\Big(\tfrac{1}{2}P_+P_-+P_+v_-+P_-v_++v_+v_-\Big)
\\
\label{eq:fieldeq3}&-\frac{\mathcal{E}_\nu}{4}\Bigl(\sqrt{\frac{\nu_+}{\nu_-
  }}\text{Tr}(v_-^2)+\sqrt{\frac{\nu_-}{\nu_+
  }}\text{Tr}(v_+^2)\Bigr)\, ,
\\
\label{eq:veq}P_\pm \doteq& -v_\pm-\frac{1}{2}\mathcal{E}_\nu\sqrt{\frac{\nu_\pm}{\nu_\mp}}v_{\mp}\, ,
\\
\label{eq:rhoeq} \partial_\pm\rho \doteq& -\chi_{2\,\pm}-\frac{1}{2}\mathcal{E}_\nu\sqrt{\frac{\nu_\pm}{\nu_\mp}}\chi_{2\,\mp}\, ,
\\
\label{eq:sigmaeq}\partial_\pm\sigma \doteq& -\chi_{1\,\pm}-\frac{1}{2}\mathcal{E}_\nu\sqrt{\frac{\nu_\pm}{\nu_\mp}}\chi_{1\,\mp}\, .
\end{align}
Let us begin with the equations in \eqref{eq:rhoeq}. From the one for $\partial_{-}\rho$ we can express $\chi_{2\,-}$ as
\begin{equation}\label{eq:initialeq}
 \chi_{2\,-}\doteq-\partial_-\rho  -\frac{1}{2}\mathcal{E}_\nu\sqrt{\frac{\nu_-}{\nu_+}}\chi_{2\,+} \, ,
\end{equation}
and substitute it back into the $\chi_{2\,-}$ contribution contained in the equation for $\partial_{+}\rho$, thus obtaining an expression for $\chi_{2\,+}$ written in terms of $\partial_\pm\rho$ and $\mathcal{E}_\nu$ only. Substituting once again this back into \eqref{eq:initialeq}, we then arrive at the rewriting
\begin{equation}
 \chi_{2\,\pm} \doteq\Big(\frac{4}{\mathcal{E}_\nu^2-4}\Big)\Big(\partial_\pm\rho-\frac{1}{2}\mathcal{E}_\nu\sqrt{\frac{\nu_\pm}{\nu_\mp}}\partial_\mp\rho\Big)\, .   
\end{equation}
Similarly, from \eqref{eq:sigmaeq} and \eqref{eq:veq} one obtains
\begin{align}
  \chi_{1\,\pm} &\doteq\Big(\frac{4}{\mathcal{E}_\nu^2-4}\Big)\Big(\partial_\pm\sigma-\frac{1}{2}\mathcal{E}_\nu\sqrt{\frac{\nu_\pm}{\nu_\mp}}\partial_\mp\sigma\Big)\, ,\\
 v_\pm &\doteq\Big(\frac{4}{\mathcal{E}_\nu^2-4}\Big)\Big(P_\pm-\frac{1}{2}\mathcal{E}_\nu\sqrt{\frac{\nu_\pm}{\nu_\mp}}P_\mp\Big)\, ,  
\end{align}
and exploiting all of the above, one finds the following two relations
\begin{align}
\nu_\pm\Big(\frac{\mathcal{E}_\nu^2-4}{4}\Big)^2\equiv& \Big(\chi_{1\,\pm}\chi_{2\,\pm}-\frac{\rho}{2}\text{Tr}(v_\pm^2)\Big)\Big(\frac{\mathcal{E}_\nu^2-4}{4}\Big)^2 
\notag \\
\doteq&+ \partial_\pm\rho\partial_\pm\sigma-\frac{\rho}{2}\text{Tr}(P_\pm^2)+\frac{\mathcal{E}_\nu^2}{4}\frac{\nu_\pm}{\nu_\mp}\Big(\partial_\mp\rho\partial_\mp\sigma-\frac{\rho}{2}\text{Tr}(P_\mp^2)\Big)
\\
&  -\frac{\mathcal{E}_\nu}{2}\sqrt{\frac{\nu_\pm}{\nu_\mp}}\Big(\partial_\pm\rho\partial_\mp\sigma+\partial_\mp\rho+\partial_\pm\sigma-\rho\text{Tr}(P_\pm P_\mp)\Big).
\notag
\end{align}
Looking at the left hand sides of these expressions, one can recognise them to become equal upon multiplication by $\nu_\mp$. In turn this implies that also the two right hand sides should be equal, and after some algebra one obtains the relations
\begin{equation}
  \frac{\nu_\pm}{\nu_\mp} \doteq \frac{I_{(\pm\pm)}}{I_{(\mp\mp)}}\, ,  
\end{equation}
where $I_{\pm\pm}$ have been defined in \eqref{eq:Jpm2}. In analogy with the PCM, we are then able to get the expressions (rewritten having already in mind the comparison with the $\mu$-frame)
\begin{align}\label{eq:auxofEnurho}
 \chi^{(\mu)}_{2\;\pm} &\doteq f(\mu)\Big(\partial_\pm\rho-\frac{1}{2}\mu\sqrt{\frac{I_{(\pm\pm)}}{I_{(\mp\mp)}}}\partial_\mp\rho\Big),\\
\label{eq:auxofEnusigma}\chi^{(\mu)}_{1\;\pm} &\doteq f(\mu)\Big(\partial_\pm\sigma-\frac{1}{2}\mu\sqrt{\frac{I_{(\pm\pm)}}{I_{(\mp\mp)}}}\partial_\mp\sigma\Big),\\
\label{eq:auxofEnuP}v^{(\mu)}_{\pm} &\doteq f(\mu)\Big(P_\pm-\frac{1}{2}\mu\sqrt{\frac{I_{(\pm\pm)}}{I_{(\mp\mp)}}}P_\mp\Big).
\end{align}
To complete the equivalence check one should at this point substitute these expressions into the field equations \cref{eq:fieldeq1,eq:fieldeq2,eq:fieldeq3}, and verify agreement with the equations arising from the action \eqref{eq:muframelagansatz}. The easiest one to consider is \eqref{eq:fieldeq1}. Substituting \cref{eq:auxofEnurho,eq:auxofEnusigma,eq:auxofEnuP} we obtain
\begin{align}
\partial_+\partial_-\rho+\partial_+\Bigl(\!f(\mu)\bigl(\!\partial_-\rho-\tfrac{1}{2}\mu\sqrt{\frac{I_{(--)}}{I_{(++)}}}\partial_+\rho\bigr)\!\!\Bigr)+\partial_-\Bigl(\!f(\mu)\bigl(\!\partial_+\rho-\tfrac{1}{2}\mu\sqrt{\frac{I_{(++)}}{I_{(--)}}}\partial_-\rho\bigr)\!\!\Bigr)\doteq 0 \,,
\end{align}
from which, after isolating the term $(1+2f(\mu))\partial_{+}\partial_{-}\rho$, one finds
\begin{align}\label{eq:exprecompared1}
\partial_+\partial_-\rho +\frac{1}{1\!+\!2f(\mu)}\!\Bigg[\!\!&+\partial_+f(\mu)\partial_-\rho +\partial_-f(\mu)\partial_+\rho-\frac{\mu f(\mu)}{2}\Bigl(\sqrt{\frac{I_{(--)}}{I_{(++)}}}\partial_+\partial_+\rho+\sqrt{\frac{I_{(++)}}{I_{(--)}}}\partial_-\partial_-\rho\Bigr)
\notag \\
&-\frac{1}{2}\Big(\sqrt{\frac{I_{(--)}}{I_{(++)}}}\partial_+(\mu f(\mu))\partial_+\rho+\sqrt{\frac{I_{(++)}}{I_{(--)}}}\partial_-(\mu f(\mu))\partial_-\rho\Big) 
\notag \\
&-\frac{\mu f(\mu)}{2}\Big( \partial_+\sqrt{\frac{I_{(--)}}{I_{(++)}}}\partial_+\rho+\partial_-\sqrt{\frac{I_{(++)}}{I_{(--)}}}\partial_-\rho\Big)\Bigg]\doteq 0\, .
\end{align}
The corresponding variation with respect to $\sigma$ of \eqref{eq:muframelagansatz},
\begin{equation}
\delta(I_{(+-)}) F(\mu)+\delta(\sqrt{I_{(++)}I_{(--)}})G(\mu)\;\dot{\approx}\; 0\, ,
\end{equation}
provides instead with, after some necessary integrations by parts,
\begin{align}
\nonumber\Bigg[-2F(\mu)\partial_+\partial_-\rho &-\partial_-\rho\partial_+f(\mu) -\partial_+\rho\partial_-f(\mu)\\
& -\frac{1}{2}\partial_+\Big(G(\mu)\sqrt{\frac{I_{(--)}}{I_{(++)}}}\partial_+\rho\Big)-\frac{1}{2}\partial_-\Big(G(\mu)\sqrt{\frac{I_{(++)}}{I_{(--)}}}\partial_-\rho\Big)\Bigg]\delta\sigma\doteq0\, , 
\end{align}
which means that
\begin{align}\label{eq:eqcompared2}
\nonumber  \partial_+\partial_-\rho+\frac{1}{2F(\mu)}\Bigg[&+\partial_-\rho\partial_+f(\mu) +\partial_+\rho\partial_-f(\mu)+\frac{1}{2}G(\mu)\Big(\sqrt{\frac{I_{(--)}}{I_{(++)}}}\partial_+\partial_+\rho+\sqrt{\frac{I_{(++)}}{I_{(--)}}}\partial_-\partial_-\rho\Big)\\
\nonumber &+\frac{1}{2}\Big(\sqrt{\frac{I_{(--)}}{I_{(++)}}}\partial_+G(\mu)\partial_+\rho+\sqrt{\frac{I_{(++)}}{I_{(--)}}}\partial_-G(\mu)\partial_-\rho\Big)\\
&+\frac{1}{2}G(\mu)\Big(\partial_+\sqrt{\frac{I_{(--)}}{I_{(++)}}}\partial_+\rho+\partial_-\sqrt{\frac{I_{(++)}}{I_{(--)}}}\partial_-\rho\Big)\Bigg]\doteq 0\, .
\end{align}
Therefore, for the two expressions \eqref{eq:exprecompared1} and \eqref{eq:eqcompared2} to agree it must be true that
\begin{align}
  \frac{1}{1+2f(\mu)}=\frac{1}{2F(\mu)} \qquad \text{and} \qquad -\frac{\mu f(\mu)}{2(1+2f(\mu))}=\frac{G(\mu)}{4F(\mu)}\, ,
\end{align}
with $F(\mu),G(\mu)$ given in \eqref{F-G-functions}, thus fixing $(c_1=\frac{1}{2},c_2=0)$ and selecting the minus sign for $g(\mu)$. With this choice, the $\mu$-frame Lagrangian \eqref{eq:muframelagansatz} takes the final form \eqref{eq:muframeLagfinal}.

Let us now turn to the field equation \eqref{eq:fieldeq3}. Substituting \cref{eq:auxofEnusigma,eq:auxofEnuP} one obtains
\begin{align}
\nonumber \partial_+\partial_-\sigma &+\partial_+\Bigl(f(\mu)\bigl(\partial_-\sigma-\frac{\mu}{2}\sqrt{\frac{I_{(--)}}{I_{(++)}}}\partial_+\sigma\bigr)\Bigr)+\partial_-\Bigl(f(\mu)\bigl(\partial_+\sigma-\frac{\mu}{2}\sqrt{\frac{I_{(++)}}{I_{(--)}}}\partial_-\sigma\bigr)\Bigr) \\
\nonumber\dot{\approx}\text{Tr}\Bigl[&-\Bigl(\frac{1}{2}+2f(\mu)+f(\mu)^2-\frac{1}{4}\mu^2 f(\mu)^2\Bigr)P_+P_-\\
&+\frac{1}{2}\Bigl(\mu f(\mu)+\frac{1}{2}\mu f(\mu)^2-\frac{1}{8}\mu^3f(\mu)^2\Bigr)\Big(\sqrt{\frac{I_{(--)}}{I_{(++)}}}P_+^2+\sqrt{\frac{I_{(++)}}{I_{(--)}}}P_-^2\Big)\Big]\, ,
\end{align}
from which, after isolating $(1+2f(\mu))\partial_{+}\partial_{-}\sigma$, one arrives at
\begin{align}
\partial_+\partial_-\sigma\!+\!\frac{1}{1\!+\!2f(\mu)}\!\Bigg[\!\!&+\partial_+f(\mu)\partial_-\sigma +\partial_-f(\mu)\partial_+\sigma -\frac{\mu f(\mu)}{2}\Big(\sqrt{\frac{I_{(--)}}{I_{(++)}}}\partial_+\partial_+\sigma \!+\!\sqrt{\frac{I_{(++)}}{I_{(--)}}}\partial_-\partial_-\sigma\Big)
\notag \\
&-\frac{\partial_+(\mu f(\mu))}{2}\sqrt{\frac{I_{(--)}}{I_{(++)}}}\partial_+\sigma-\frac{\partial_-(\mu f(\mu))}{2}\sqrt{\frac{I_{(++)}}{I_{(--)}}}\partial_-\sigma\Bigg]
\notag \\
\dot{\approx}\frac{1}{1\!+\!2f(\mu)}\!\text{Tr}\!\Bigg[\!\!&-\!\Bigl(\frac{1}{2}+2f(\mu)+f(\mu)^2-\frac{1}{4}\mu^2 f(\mu)^2\Bigr)P_+P_-
\notag \\
&+\frac{1}{2}\Bigl(\mu f(\mu)\!+\!\frac{1}{2}\mu f(\mu)^2\!-\!\frac{1}{8}\mu^3f(\mu)^2\Bigr)\!\Big(\sqrt{\frac{I_{(--)}}{I_{(++)}}}P_+^2\!+\!\sqrt{\frac{I_{(++)}}{I_{(--)}}}P_-^2\Big)\!\!\Bigg].
\end{align}
By then noticing that 
\begin{equation}
\frac{1}{1+2f(\mu)}\Big(\frac{1}{2}+2f(\mu)+f(\mu)^2-\frac{1}{4}\mu^2 f(\mu)^2\Big)=\frac{1}{2}\, , 
\end{equation}
one arrives at the expression
\begin{align}\label{eq:checksigma}
&\partial_+\partial_-\sigma\!+\!\frac{1}{1\!+\!2f(\mu)}\!\Bigg[\!\!+\!\partial_+f(\mu)\partial_-\sigma \!+\! \partial_-f(\mu)\partial_+\sigma \!-\! \frac{\mu f(\mu)}{2}\Big(\sqrt{\frac{I_{(--)}}{I_{(++)}}}\partial_+\partial_+\sigma \!+\! \sqrt{\frac{I_{(++)}}{I_{(--)}}}\partial_-\partial_-\sigma\Big)
\notag\\
&\qquad \qquad \qquad \qquad \qquad -\frac{\partial_+(\mu f(\mu))}{2}\sqrt{\frac{I_{(--)}}{I_{(++)}}}\partial_+\sigma-\frac{\partial_-(\mu f(\mu))}{2}\sqrt{\frac{I_{(++)}}{I_{(--)}}}\partial_-\sigma\Bigg]
\\
&\dot{\approx}-\tfrac{1}{2}\text{Tr}\!\Bigg[P_+P_- -\frac{1}{1+2f(\mu)}\Bigl(\mu f(\mu)\!+\!\frac{1}{2}\mu f(\mu)^2\!-\!\frac{1}{8}\mu^3f(\mu)^2\Bigr)\Big(\sqrt{\frac{I_{(--)}}{I_{(++)}}}P_+^2 \!+\! \sqrt{\frac{I_{(++)}}{I_{(--)}}}P_-^2\Big)\Bigg]\, .
\notag
\end{align}
On the other hand, from the variation of the $\mu$-frame action one obtains
\begin{align}
\partial_+\partial_-\sigma&+\frac{1}{2}\text{Tr}(P_+P_-)+\frac{1}{2F(\mu)}(\partial_+f(\mu)\partial_-\sigma+\partial_-f(\mu)\partial_+\sigma) 
\notag \\
&+\frac{1}{4F(\mu)}\Bigg[+G(\mu)\Bigl(\sqrt{\frac{I_{(--)}}{I_{(++)}}}\partial_+\partial_+\sigma+\sqrt{\frac{I_{(++)}}{I_{(--)}}}\partial_-\partial_-\sigma\Bigr)
\notag \\
&\qquad \quad \quad \,\,\,\,  +\partial_+G(\mu)\sqrt{\frac{I_{(--)}}{I_{(++)}}}\partial_+\sigma+\partial_-G(\mu)\sqrt{\frac{I_{(++)}}{I_{(--)}}}\partial_-\sigma
\notag \\
&\qquad \quad \quad \,\,\,\,+G(\mu)\Bigl(\partial_+\sqrt{\frac{I_{(--)}}{I_{(++)}}}\partial_+\sigma+\partial_-\sqrt{\frac{I_{(++)}}{I_{(--)}}}\partial_-\sigma\Bigr)
\notag \\
& \qquad \quad \quad \,\,\,\,-\frac{1}{2}G(\mu)\sqrt{\frac{I_{(--)}}{I_{(++)}}}\text{Tr}(P_+^2)-\frac{1}{2}G(\mu)\sqrt{\frac{I_{(++)}}{I_{(--)}}}\text{Tr}(P_-^2)\Bigg]\dot{\approx}0\, ,
\end{align}
and the several coefficients match the ones in \eqref{eq:checksigma}. Finally turning to the last remaining equation of motion \eqref{eq:fieldeq2}, upon substituting \eqref{eq:auxofEnuP} one finds
\begin{align}
\nonumber  \partial_+\rho\Big(P_- &-\frac{\mu f(\mu)}{1+2f(\mu)}\sqrt{\frac{I_{(--)}}{I_{(++)}}}P_+\Big) +\partial_-\rho\Big(P_+-\frac{\mu f(\mu)}{1+2f(\mu)}\sqrt{\frac{I_{(++)}}{I_{(--)}}}P_-\Big) \\
\nonumber&+\frac{\rho}{1+2f(\mu)}\Bigg(+D_+\Big((1+2f(\mu)) P_--\mu f(\mu)\sqrt{\frac{I_{(--)}}{I_{(++)}}}P_+\Big)\\
&\qquad \qquad \qquad \,\,\, +D_+\Big((1+2f(\mu)) P_+-\mu f(\mu)\sqrt{\frac{I_{(++)}}{I_{(--)}}}P_-\Big)\Bigg)\dot{\approx}0\, ,
\end{align}
which should be compared with the equations stemming from varying \eqref{eq:muframeLagfinal} with respect to the symmetry-group degrees of freedom,
\begin{align}
\nonumber  \rho(D_+P_- &+D_-P_+)+\partial_+\rho\Big(P_-+\frac{G(\mu)}{2F(\mu)}\sqrt{\frac{I_{(--)}}{I_{(++)}}}P_+\Big) +\partial_-\rho\Big(P_++\frac{G(\mu)}{2F(\mu)}\sqrt{\frac{I_{(++)}}{I_{(--)}}}P_-\Big)
\\
&+\frac{\rho}{F(\mu)}\!\Bigg[\!\!+\partial_+F(\mu)P_-+\partial_-F(\mu)P_++\frac{1}{2}\sqrt{\frac{I_{(++)}}{I_{(--)}}}\partial_-G(\mu)P_-+\frac{1}{2}\sqrt{\frac{I_{(--)}}{I_{(++)}}}\partial_+G(\mu)P_+
\notag \\
&\qquad \quad \,\,\,\,  +\frac{1}{2}\partial_-\sqrt{\frac{I_{(++)}}{I_{(--)}}}G(\mu)P_-+\frac{1}{2}\partial_+\sqrt{\frac{I_{(--)}}{I_{(++)}}}G(\mu)P_+\Bigg]\dot{\approx}0\, ,
\end{align}
again finding perfect match.

\section{Linearisation of the TEGR action}\label{sec:linearisation}
In this appendix we scrutinise the linearisation of the theory \eqref{eq:SEHaux} around flat Minkowski spacetime, examining its relation to the linearised self-dual formulation of \cite{Henneaux:2004jw,Bunster:2012km}. 

We begin by considering the following vierbein
\be\label{eq:vierbeinlin}
E_M{}^A=\delta_M{}^A+ \kappa \,e_M{}^A\, ,
\ee
where $\kappa$ is a small expansion parameter and we can ignore the distinction between flat and curved indices. In particular, we will adopt the convenient notation
\begin{equation}
 e_{AB}=\delta_{A}{}^{M} e_{M\,B},\quad \partial_A=\delta_{A}{}^{M}\partial_M,  
\end{equation}
where $ e_{AB}$ has in principle no symmetry in $\{AB\}$. The anholonomicity coefficients become
\begin{equation}\label{eq:linearOmega}
    \Omega_{AB}{}^C=2\kappa\,\partial_{[A}e_{B]}{}^C\, , \qquad   \Omega_{AC}{}^C=\kappa\,(\partial_A e_C{}^C-\partial_C e_A{}^C)\, ,
\end{equation}
and we can introduce an analogue linearised ansatz for the auxiliary field $Y_{AB}{}^C\rightarrow \kappa\, y_{AB}{}^C$, with the same symmetry properties as \eqref{eq:linearOmega}. It is then a straightforward exercise to show that, to order $\mathcal{O}(\kappa^2)$, the action \eqref{eq:SEHaux} takes the form 
\begin{align}\label{eq:SEHauxlin}
S_{\text{EH-lin}}\!=\!\frac{1}{8}\!\!\int\!\!\! d^4 x \Big[\!\!&+2\,\partial^A e^{BC}\Big(\partial_A e_{BC}+ \partial_A e_{CB}-\partial_B e_{AC}-\partial_C e_{AB}
-\partial_B e_{CA}+\partial_C e_{BA}\Big)
\notag \\
\nonumber 
&-4\Big(-2\partial_A e_B{}^B\partial^C e^A{}_C+\partial_A e_B{}^B\partial^A e^C{}_C+\partial_B e_A{}^B\partial^C e^A{}_C\Big)\\[2pt]
\nonumber &+8\Big(y^{ABC}\partial_A e_B{}^C+y^{ABC}\partial_A e_{CB}-y^{ABC}\partial_C e_{AB}-2y_{AB}{}^B(\partial_A e_C{}^C-\partial_C e^{AC})\Big)\\[2pt]
&+2(y^{ABC}y_{ABC}+2y^{ABC}y_{ACB}-4y_{AB}{}^By^{AC}{}_C)\Big]  \, .
\end{align}

One might now wonder how is this expression related to the linearised action for gravity in the Fierz-Pauli form \cite{Henneaux:2004jw}, given that we know the latter is automatically invariant under duality transformations. Let us therefore compare the first two lines of \eqref{eq:SEHauxlin} with the Fierz-Pauli action of \cite{Henneaux:2004jw}, which we rewrite here for convenience
\begin{align}\label{eq:FPaction}
\!\!\!\!S_{\text{FP}}\!=\!-\tfrac{1}{4}\!\!\int\!\! d^4x\!\Big[\!\partial^R h^{MN}\partial_R h_{MN}\!-\!2\partial_M h^{MN}\partial_R h^R{}_N\!+\!2\partial^M h^R{}_R\partial^N h_{MN}\!-\!\partial^M h^R{}_R\partial_M h^Q{}_Q\!\Big] .  
\end{align}
The field $h_{MN}=h_{NM}$ is the spin-2 metric perturbation around a Minkowski vacuum,
\be
G_{MN}=\eta_{MN}+ \lambda \, h_{MN}\, ,
\ee
and a quick comparison with \eqref{eq:vierbeinlin} shows its relation to the vierbein perturbation as \footnote{In general one might consider a priori different small-parameter expansions of the vielbeine and metric
\begin{equation}\label{eq:metricandvierbeinexp}
E_{M}{}^{A}=\delta_{M}{}^{A}+\kappa \, e_{M}{}^{A} \qquad \text{and} \qquad 
G_{MN}=\eta_{MN}+ \lambda \, h_{MN} \,\, ,
\end{equation}
but the relation $G_{MN}=\eta_{AB}E_{M}{}^{A}E_{N}{}^{B}$, which translates into the condition
\begin{equation}\label{h-e-relation}
 G_{MN}=\eta_{MN}+\lambda h_{MN}=\eta_{MN}+\kappa(e_{MN}+e_{NM})+\kappa^2 e_M^P e_{PN}\, ,   
\end{equation}
in fact imposes that $\kappa = \kappa(\lambda)$ in a quite restricted fashion. Indeed, agreement of \eqref{h-e-relation} at $\lambda=0$ enforces that $\kappa(0)=0$, such that a Taylor expansion of $\kappa(\lambda)$ on the r.h.s leads to
\begin{equation}
   \eta_{MN}+\lambda h_{MN}=\eta_{MN}+\kappa^\prime (0)(e_{MN}+e_{NM}) \lambda+\mathcal{O}(\lambda^2)\, ,
\end{equation}
such that even for a general relation between $\kappa$ and $\lambda$, the metric perturbation $h_{MN}$ can at most be proportional to the symmetric combination of vielbeine perturbation as in \eqref{metric-pertubation}.
}
\begin{equation}\label{metric-pertubation}
    h_{MN}=e_{MN}+e_{NM}\, .
\end{equation}

Expressing \eqref{eq:FPaction} in terms of this new variable, one finds
\begin{align}\label{eq:FPactionh}
  S_{\text{FP}}\!=\!-\tfrac{1}{2}\!\!\int \!\!\!d^4x\!\Big[+&\partial^R e^{MN}(\partial_R e_{MN}+\partial_R e_{NM})-\partial_M e^{MN}\partial_R e^R{}_N-2\partial_M e^{MN} \partial_R e_N{}^R\\
  -&\partial_M e^{NM}\partial_R e_N{}^R+2\partial^M e^R{}_R\partial^N e_{MN}+2\partial^M e^R{}_R\partial^N e_{NM}-2\partial^M e^R{}_R\partial_M e^S{}_S\Big]\,.
  \notag 
\end{align}
We remark that due to the expansion around Minkowski vacuum, we can ignore the distinction between flat and curved indices, and that all indices are raised and lowered with the Minkowski metric $\eta_{MN}$, $\eta_{AB}$; had another vacuum been chosen, the analysis should have been carried out more carefully. 

Taking into account an overall factor of $\frac{1}{2}$ in order to match normalisations, the action \eqref{eq:FPactionh} can be shown to be equivalent, via several integrations by parts, to minus the first two lines in \eqref{eq:SEHauxlin}, corresponding to the linearised Einstein-Hilbert action. To check this, we start by rearranging the two actions so as to highlight their common part
\begin{equation}
S_{\text{FP}} = -\tfrac{1}{2} (S_{0} + S_{\text{FP}}^{\text{extra}}) 
\qquad \text{and} \qquad 
S_{\text{EH-lin}} = -\tfrac{1}{4}(S_{0} + S_{\text{EH-lin}}^{\text{extra}})
\end{equation}
where
\begin{equation}
\!\!\!S_{0}\!\!=\!\!\!\!\int\!\! d^{4}x \Big[\! \partial^{R}e^{MN}\partial_{R}(\!e_{MN}+e_{NM}\!)-\partial_{R}e_{N}{}^{R}\partial_{M}e^{NM}+2\partial^{M}e_{P}{}^{P}\partial^{N}e_{MN}-2\partial^{M}e_{P}{}^{P}\partial_{M}e_{Q}{}^{Q} \!\Big]
\end{equation}
and the extra contributions read
\begin{align}
S_{\text{FP}}^{\text{extra}}\!\!&=\!\!\!\int \!\!\! d^{4}x [-\partial_{R}e^{R}{}_{N}\partial_{M}(\!e^{MN}+e^{NM}\!)\!-\!\partial_{R}e_{N}{}^{R}\partial_{M}e^{MN}\!+\!2\partial^{M}e_{P}{}^{P}\partial^{N}e_{NM}]
\\
S_{\text{EH-lin}}^{\text{extra}}\!\!&=\!\!\!\int \!\!\! d^{4}x[-\partial^{A}e^{BC}\partial_{B}(\!e_{AC}+e_{CA}\!)\!-\!\partial^{A}e^{BC}\partial_{C}(\!e_{AB}\!-\!e_{BA}\!)\!-\!\partial_{B}e_{A}{}^{B}\partial_{C}e^{AC}\!+\!2\partial_{A}e_{B}{}^{B}\partial^{C}e^{A}{}_{C}]
\notag
\end{align}
At this point it is not hard to check that also the latter two are the same, at least up to boundary terms. Indeed, integrating by parts twice the first two terms in $S_{\text{EH-lin}}^{\text{extra}}$ and adding and subtracting a term $2\partial^{A}e_{B}{}^{B}\partial^{C}e_{CA}$ one finds, discarding boundary terms,
\begin{align}
S_{\text{EH-lin}}^{\text{extra}} &= \int d^{4}x [-\partial_{B}e^{BC}\partial^{A}(e_{AC}+e_{CA})-\partial_{C}e^{BC}\partial^{A}(e_{AB}-e_{BA})+
\notag \\
& \qquad \qquad  -\partial_{B}e_{A}{}^{B}\partial_{C}e^{AC}+2\partial_{A}e_{B}{}^{B}\partial^{C}e^{A}{}_{C}+ 2\partial^{A}e_{B}{}^{B}\partial^{C}e_{CA}-2\partial^{A}e_{B}{}^{B}\partial^{C}e_{CA}] 
\notag \\
&=   \int d^{4}x [-\partial_{B}e^{BC}\partial^{A}(e_{AC}+e_{CA})-\partial_{C}e^{BC}\partial^{A}e_{AB}
\notag \\
& \qquad \qquad
+2\partial_{A}e_{B}{}^{B}\partial^{C}e^{A}{}_{C}+ 2\partial^{A}e_{B}{}^{B}\partial^{C}e_{CA}-2\partial^{A}e_{B}{}^{B}\partial^{C}e_{CA}] 
\notag \\
&= S_{\text{FP}}^{\text{extra}} +2 \int d^{4}x \, \partial^{A}e_{B}{}^{B}\partial^{C}(e_{AC}-e_{CA})
\notag \\
&= S_{\text{FP}}^{\text{extra}} 
\end{align}
after integrating by parts in the last step and using symmetry of partial derivatives.
\section{Transformation properties of $\mathcal{K}[E]$ under infinitesimal diffeomorphisms}\label{sec:appendixsymmetries}
Under the action of $D=4$ diffeomorhpisms, the vierbein determinant transforms as a scalar density of weight +1
\begin{equation}
    \delta_\xi E=\xi^P\partial_PE+E\partial_P\xi^P=\partial_P\big(\xi^P E\big)\, ,
\end{equation}
while the connection $W_{AB}{}^C$ does not transform properly as a scalar,
\begin{equation}
    \delta_\xi W_{AB}{}^C=\xi^P\partial_PW_{AB}{}^C+\Delta_{AB}{}^C,
\end{equation}
due to the additional piece
\begin{equation}\label{eq:anomalousterm}
 \Delta_{AB}{}{^C}\equiv   E_A{}^ME_B{}^N E_P{}^C\partial_N\partial_M\xi^P.
\end{equation}
Notice that, on the contrary, the anholonomicity coefficients, transform as a proper scalar
\begin{equation}
  \delta_\xi\Omega_{AB}{}^C=\xi^P\partial_P\Omega_{AB}{}^C\, ,  
\end{equation}
due to their antisymmetry in the first two indices, which  kills the anomalous term \eqref{eq:anomalousterm}. 

Thus, the variation under diffeomorphisms of the object $t_{AB}$ defined in \eqref{eq:finalt} is 
\begin{align}
\nonumber&\delta_\xi t_{AB}=\delta_\xi T_{AB}-\delta_\xi\big(E[S_\mu, A^\mu]_{AB}\big) =\partial_P\Big(\xi^P T_{AB}\Big)+\frac{1}{8}
\delta_\xi\Big(E\big( W_{DAC}W^{D}{}_{B}{}^{C}-W_{DCA}W^{DC}{}_B\big)\Big)  \\[3pt]
\nonumber&=\partial_P\Big(\xi^P T_{AB}-E[S_\mu, A^\mu]_{AB}\Big)+\frac{E}{8}(\Delta_{DA}{}^C W_{FB}{}^G+W_{DA}{}^C\Delta_{FB}{}^G)\eta^{DF}\eta_{CG}\\
\nonumber&\hspace{5.4cm}-\frac{E}{8}(\Delta_{DC}{}^FW_{IH}{}^G+W_{DC}{}^F\Delta_{IH}{}^G)\eta_{AF}\eta_{BG}\eta^{DI}\eta^{CH}\\[3pt]
&=\partial_P(\xi^P t_{AB})-\frac{E}{8}\tilde{\delta}_\xi[S_\mu, A^\mu]_{AB}\, ,
\end{align}
where we have defined
\begin{equation}
\begin{aligned}
\tilde{\delta}_\xi[S_\mu, A^\mu]_{AB} =& 
-(\Delta_{DA}{}^C W_{FB}{}^G+W_{DA}{}^C\Delta_{FB}{}^G)\eta^{DF}\eta_{CG} 
\\
&+(\Delta_{DCA}W_{IHB}+W_{DCA}\Delta_{IHB})\eta^{DI}\eta^{CH}.
\end{aligned}
\end{equation}

The total variation of the $\mathcal{K}[E]$ object will then be
\begin{align}\label{eq:diffvariationofK}
\nonumber& \delta_\xi \mathcal{K}[E]=  (2\eta^{AB}\eta^{CD}-\eta^{AC}\eta^{BD})\delta_\xi(t_{AC}t_{BD})\\[3pt]
&\nonumber=(2\eta^{AB}\eta^{CD}-\eta^{AC}\eta^{BD})\big(\delta_\xi(t_{AC})t_{BD}+\delta_\xi(t_{BD})t_{AC}\big) \\[3pt]
 \nonumber &=(2\eta^{AB}\eta^{CD}-\eta^{AC}\eta^{BD})\Big(\partial_P(\xi^P t_{AC})t_{BD}+t_{AC}\partial_P(\xi^Pt_{BD})\Big)\\
 &\nonumber \,\,\,\,\,\,+\frac{E}{8}\Bigg(\!\!(2\eta^{AB}\eta^{CD}-\eta^{AC}\eta^{BD})\, t_{BD}\,\tilde{\delta}_\xi[S_\mu, A^\mu]_{AC}+(2\eta^{AB}\eta^{CD}-\eta^{AC}\eta^{BD})\, t_{AC}\,\tilde{\delta}_\xi[S_\mu, A^\mu]_{BD}\!\!\Bigg)\, \\[3pt]
 \nonumber&=(2\eta^{AB}\eta^{CD}-\eta^{AC}\eta^{BD})\Big[\partial_P\big(\xi^P t_{AC}t_{BD}\big)+t_{AC}t_{BD}\partial_P\xi^P\\
 \nonumber & \hspace{7.22cm}+\frac{E}{8}\,\Big( t_{AC}\,\tilde{\delta}_\xi[S_\mu, A^\mu]_{BD}+t_{BD}\,\tilde{\delta}_\xi[S_\mu, A^\mu]_{AC}\Big)\Big]\\[3pt]
\nonumber&= \partial_P\big(\xi^P \mathcal{K}[E]\big)+\mathcal{K}[E]\partial_P\xi^P\\
\nonumber&\hspace{2.75cm}+(2\eta^{AB}\eta^{CD}-\eta^{AC}\eta^{BD})\frac{E}{8}\, \big(t_{AC}\,\tilde{\delta}_\xi[S_\mu, A^\mu]_{BD}+t_{BD}\,\tilde{\delta}_\xi[S_\mu, A^\mu]_{AC}\big)\\[3pt]
\nonumber &=\partial_P\big(\xi^P \mathcal{K}[E]\big)+\mathcal{K}[E]\partial_P\xi^P+2\eta^{AB}\eta^{CD}\frac{E}{8}\, \big(t_{AC}\,\tilde{\delta}_\xi[S_\mu, A^\mu]_{BD}+t_{BD}\,\tilde{\delta}_\xi[S_\mu, A^\mu]_{AC}\big)\, \\[3pt]
&=\partial_P\big(\xi^P \mathcal{K}[E]\big)+\mathcal{K}[E]\partial_P\xi^P+4\eta^{AB}\eta^{CD}\frac{E}{8}\, t_{AC}\,\tilde{\delta}_\xi[S_\mu, A^\mu]_{BD},
\end{align}

where in the second to last line we have used the fact that $\eta^{AB}\tilde{\delta}_\xi[S_\mu, A^\mu]_{AB}$ is zero, as can be explicitly checked.
The first of the three terms transforms appropriately as a scalar density of weight +1: unfortunately, two obstructions embodied in the second and third terms respectively, prevent full diffeomorphisms invariance. The first one is due to the scaling of the operator with $E^2$, rather than with $E$, and leads to its transformation behaviour as a scalar density of weight +2, rather than +1. The second is the double derivative term acting on the arbitrary diffeomorphism parameter $\xi^P$, due to the lack of antisymmetry of the components of the second term in \eqref{eq:finalt}. 

While it is immediately obvious how these transformation properties spoil diffeomorphism invariance of \eqref{eq:fullmuframeLag}, they do so also with the deformed action \eqref{eq:finalLagnuframe}. This is can be inferred either by assessing that $Z_{AB}{}^C$ and $Y_{AB}{}^C$ share the same transformation properties of $W_{AB}{}^C$ and $\Omega_{AB}{}^C$, or by realising that, solving the field equations in a power series expansion in terms of the coupling constant $\epsilon$, we get 
\begin{align}
\epsilon \nu_4 = \epsilon\mathcal{K}[E] +\mathcal{O}(\epsilon^p), \quad p>1 . 
\end{align}
In particular, as the order in the expansion increases, so do the powers of $\mathcal{K}$ and the ``pathological" behaviour under diffeomorphisms becomes progressively worse.

It is important to notice that the theory is still invariant under the subset of diffeomorphisms satisfying the divergence-free condition,
\begin{equation}
    \partial_Q\xi^Q=0\, ,
\end{equation}
which in turn restricts the full set of diffeomorphisms to the transverse ones, and whose $\xi^P(x)$ can furthermore only be at most linear functions of the original coordinates $x^M$ 
Therefore a SL(4) symmetry is preserved.

One attempt to save at least the correct number of degrees of freedom and not to incur into the appearance of ghost fields, would be to show the invariance of the deformed theory under ``deformed" diffeomorphism, in the spirit of the BRST cohomological approach \cite{Henneaux:1997bm,Barnich:1995ap}. More specifically, one could try to modify diffeomorphisms
\begin{equation}
  \delta_\xi = \overset{(0)}{\delta_\xi}+\epsilon \overset{(1)}{\delta_\xi}+\mathcal{O}(\epsilon^2),
\end{equation}
in such a way that
\begin{equation}
 \delta_\xi\mathcal{L}=\overset{(0)}{\delta_\xi}\overset{(0)}{\mathcal{L}} +\epsilon \big(\overset{(1)}{\delta_\xi}\overset{(0)}{\mathcal{L}}+\overset{(0)}{\delta_\xi}\overset{(1)}{\mathcal{L}}\big) +\mathcal{O}(\epsilon^2),
\end{equation}
and therefore demanding that
\begin{equation}
   \overset{(1)}{\delta_\xi}\overset{(0)}{\mathcal{L}}+\overset{(0)}{\delta_\xi}\overset{(1)}{\mathcal{L}}=0\,. 
\end{equation}
$\overset{(0)}{\delta_\xi}\overset{(1)}{\mathcal{L}}$ is precisely what is computed in \eqref{eq:diffvariationofK}. We can then employ an ansatz
\begin{equation}
 \overset{(1)}{\delta_\xi}E_M{}^A=f_M{}^A [\xi, E, \partial E]\, ,   
\end{equation}
such that
\begin{equation}
\begin{aligned}
 \overset{(1)}{\delta_\xi}E=&+E E_A{}^M f_M{}^A,
 \\
 \overset{(1)}{\delta_\xi}W_{AB}{}^C=&-W_{DB}{}^C E_A{}^R f_R{}^D-W_{AD}{}^C E_B{}^R f_R{}^D +E_A{}^ME_B{}^N\partial_M f_N{}^C,
 \\
 \overset{(1)}{\delta_\xi}\Omega_{AB}{}^C=&+2E_{[A}{}^R\Omega_{B]D}{}^C f_R{}^D+2E_{[A}{}^ME_{B]}{}^N\partial_M f_N{}^C .  
\end{aligned}
\end{equation}

Therefore, by varying 
\begin{equation}
 \overset{(0)}{\mathcal{L}}\equiv\mathcal{L}_{EH}= E\overset{(0)}{L},    
\end{equation}
with 
\begin{align}
 \overset{(0)}{L}\equiv  \frac{1}{8}\left(\Omega^{ABC}\Omega_{ABC}+2\Omega^{ABC}\Omega_{ACB}-4\Omega_{AB}{}^B\Omega^{AC}{}_C \right),
\end{align}
we obtain, piecewise,
\begin{align}
 \overset{(1)}{\delta_\xi}E\overset{(0)}{L}&= E_A{}^M f_M{}^A \overset{(0)}{\mathcal{L}},\\
E\overset{(1)}{\delta_\xi}(\Omega_{ABC}\Omega^{ABC})&=4E\Omega^{ABC} (E_A{}^Rf_R{}^D\Omega_{BD}{}^C+E_A{}^ME_B{}^N\partial_Mf_N{}^C)\, ,\\
2E\overset{(1)}{\delta_\xi}(\Omega_{ABC}\Omega^{ACB})&=4E\Omega^{ACB}(E_A{}^Rf_R{}^D\Omega_{BD}{}^C-E_B{}^Rf_R{}^D\Omega_{AD}{}^C+2E_{[A}{}^ME_{B]}{}^N\partial_Mf_N{}^C)\, ,\\
-4E\overset{(1)}{\delta_\xi}(\Omega_{AC}{}^C\Omega^{AD}{}_D)&=-8E\Omega^{AF}{}_F(E_C{}^Rf_R{}^D\Omega_{AD}{}^C+E_A{}^R f_R{}^D\Omega_{DC}{}^C+2E_A{}^NE_C{}^M\partial_{[M}f_{N]}{}^C).
\end{align}

This means that 
\begin{align}
 \nonumber \overset{(1)}{\delta_\xi} \overset{(0)}{\mathcal{L}} =&  E_A{}^M f_M{}^A \overset{(0)}{\mathcal{L}}+\frac{1}{2}E\Omega^{AB}{}_{C} (E_A{}^Rf_R{}^D\Omega_{BD}{}^C+E_A{}^ME_B{}^N\partial_Mf_N{}^C)\\
\nonumber& +\frac{1}{2}E\Omega^{ACB}(E_A{}^Rf_R{}^D\Omega_{BDC}-E_B{}^Rf_R{}^D\Omega_{ADC}+2E_{[A}{}^ME_{B]}{}^N\partial_Mf_{\;C})\\
\nonumber&-E\Omega^{AF}{}_F(E_C{}^Rf_R{}^D\Omega_{AD}{}^C+E_A{}^R f_R{}^D\Omega_{DC}{}^C+2E_A{}^NE_C{}^M\partial_{[M}f_{N]}{}^C)\\[3pt]
\nonumber=&E_A{}^M f_M{}^A \overset{(0)}{\mathcal{L}}+EE_{[A}{}^ME_{B]}{}^N\partial_Mf_N{}^C (\frac{1}{2}\Omega^{AB}{}_{C}+\Omega^{A}{}_{C}{}^{B}-2\Omega^{AF}{}_F\delta_C^B)\\
\nonumber&+E\Omega^{ABC} E_A{}^Rf_R{}^D\Omega_{BDC}-\frac{1}{2}E\Omega^{ACB}E_B{}^Rf_R{}^D\Omega_{ADC}\\
&-E\Omega^{AF}{}_F(E_C{}^Rf_R{}^D\Omega_{AD}{}^C+E_A{}^R f_R{}^D\Omega_{DC}{}^C)
.
\end{align}
Upon a suitable choice of $f_M{}^A$ and possibly upon several integration by parts, this term could in principle conspire to cancel the last two terms in \eqref{eq:diffvariationofK}. We have not been able to find such a choice. In particular, trying with the ansatz $\xi_M{}^A=\overset{(0)}{\mathcal{L}}E_Q{}^A\partial_M\xi^Q$, the $\partial_M\partial_N\xi^Q$ term is automatically killed by antisymmetry.

\section{Geometric objects in the KK framework}\label{sec:appendixgeometry}
In this section, we collect several useful explicit expression for the geometrical objects, once dimensionally reduced to $D=2$. We provide all expressions directly in the conformal gauge for the zweibein.
\subsection{Einstein tensor}
Upon KK-ansatz \eqref{eq:vierbein1}, the vierbein  takes the form
\begin{equation}
    E_M{}^A=\begin{pmatrix}
        e^{\sigma}\delta_\mu{}^\alpha&0\\
        0& \rho^{\frac{1}{2}}\mathcal{V}_m{}^a
    \end{pmatrix},\quad E_A{}^M=\begin{pmatrix}
        e^{-\sigma}\delta_\alpha{}^\mu&0\\
        0& \rho^{-\frac{1}{2}}\mathcal{V}_a{}^m
    \end{pmatrix}\,.
\end{equation}

The Christoffel symbols $ \Gamma_{MN}{}^P$ have only non-null components
\begin{align}
\Gamma_{\mu\nu}{}^\rho&=2\delta_{(\mu}^{\rho}\partial_{\nu)}\sigma-\eta_{\mu\nu}\partial^\rho\sigma\, ,\\[5pt]
\nonumber\Gamma_{mn}{}^\rho&=-\frac{1}{2}e^{-2\sigma}(\rho\partial^\rho\text{ln}\rho \mathcal{V}_m{}^a\mathcal{V}_n{}^a+2\rho\mathcal{V}^a_{(m}\partial^\rho\mathcal{V}_{n)}{}^a)
\\
&=-\frac{1}{2}e^{-2\sigma}(\rho\partial^\rho\text{ln}\rho \mathcal{V}_m{}^a\mathcal{V}_n{}^a+2\rho\mathcal{V}^a_{(m}\mathcal{V}_{n)}{}^b\mathcal{J}^\rho{}_{ab})\notag\\
&=-\frac{1}{2}e^{-2\sigma}(\rho\partial^\rho\text{ln}\rho \mathcal{V}_m{}^a\mathcal{V}_n{}^a+2\rho\mathcal{V}^a_{m}\mathcal{V}_{n}{}^bP^\rho{}_{ab})\, ,\\
 \Gamma_{\mu m}{}^n&=\frac{1}{2}(\partial_\mu\text{ln}\rho\delta^n_m+\mathcal{V}_a{}^n\mathcal{V}_m{}^b\mathcal{J}_\mu{}^a{}_b+\mathcal{V}_a{}^n\partial_\mu\mathcal{V}_m{}^a)=\frac{1}{2}(\partial_\mu\text{ln}\rho\delta^n_m+2\mathcal{V}_a{}^n\mathcal{V}_m{}^b P_\mu{}^a{}_b)\,.
\end{align}
The Riemann tensor
\begin{equation}
    R^R{}_{SMN}=\partial_M\Gamma_{NS}{}^R-\partial_N\Gamma_{MS}{}^R+\Gamma_{ML}{}^R\Gamma_{NS}{}^L-\Gamma_{NL}{}^R\Gamma_{MS}{}^L\,
\end{equation}
has components
\begin{equation}
R^R{}_{SMN}=(R^\alpha{}_{\beta\mu\nu}=\mathring{R}^\alpha{}_{\beta\mu\nu}, R^p{}_{\beta m \nu},R^p{}_{\beta \nu m}=-R^p{}_{\beta m \nu}, R^\rho{}_{m \mu n},R^\rho{}_{m n \mu}=-R^\rho{}_{m \mu n}, R^p{}_{lmn},...)\,,   
\end{equation}
where the dots denote terms that do not contribute to the Ricci tensor, and with
\begin{align}
R^\alpha{}_{\beta\mu\nu}&=\mathring{R}^\alpha{}_{\beta\mu\nu},\\
 R^p{}_{\beta m \nu}&=-\partial_\nu\Gamma_{m\beta}{}^p+\Gamma_{m\lambda}{}^p\Gamma_{\nu\beta}{}^\lambda -\Gamma_{\nu l}{}^p\Gamma_{m \beta}{}^l,\\
 R^\rho{}_{m \mu n}&=\partial_\mu\Gamma_{mn}{}^\rho+\Gamma_{\mu\nu}{}^\rho\Gamma_{nm}{}^\nu-\Gamma_{np}{}^\rho\Gamma_{\mu m}{}^p,\\
 R^p{}_{lmn}&=\Gamma_{m\mu}{}^p\Gamma_{nl}{}^\mu-\Gamma_{n\mu}{}^p\Gamma_{ ml}{}^\mu.
\end{align}
Here, by $\mathring{R}^\alpha{}_{\beta\mu\nu}$ we mean that the external components correspond to the ones obtained starting from the external 2-dimensional metric. The external and internal components of the Ricci tensor can be shown to be
\begin{align}
  R_{\mu\nu}   =&-\partial_\alpha\partial^\alpha\sigma\eta_{\mu\nu}-\partial_\nu\partial_\mu \text{ln}\rho +2\partial_{(\mu} \text{ln}\rho\partial_{\nu)}\sigma -\partial_\lambda\text{ln}\rho\partial^\lambda\sigma\;\eta_{\mu\nu}-\frac{1}{2}\partial_\mu \text{ln}\rho\;\partial_\nu \text{ln}\rho-P_\mu{}^{ab}P_\nu{}_{ab}\, ,\\[5pt]
 \nonumber R_{mn}=&\,\partial_\mu\Gamma_{mn}{}^\mu+\Gamma_{\mu\nu}{}^\mu\Gamma_{nm}{}^\nu-2\Gamma_{p(m}{}^\mu\Gamma_{n)\mu}{}^p+\Gamma_{p\mu}{}^p\Gamma_{nm}{}^\mu\\
 \nonumber &=e^{-2\sigma}\rho\Big[-\frac{1}{2}\partial_\mu\partial^\mu\text{ln}\rho\cV_{m}{}^a\cV_n{}^a-\cV_m{}^a\cV_n{}^b(\partial_\mu P^\mu{}_{ab}+Q_\mu{}^c{}_aP^\mu_{bc}+Q_\mu{}^c{}_bP^\mu_{ac})\\
 &\hspace{5cm}-\frac{1}{2}\partial_\mu\text{ln}\rho\partial^\mu\text{ln}\rho\cV_m{}^a\cV_n{}^a-\cV_m{}^a\cV_n{}^b\partial_\mu\text{ln}\rho P^\mu{}_{ab}\Big] .
\end{align}

Therefore, the Ricci scalar is readily obtained as 
\begin{align}
    \nonumber R=&e^{-2\sigma}\eta^{\mu\nu}R_{\mu\nu}+\rho^{-1}\cV^{m}{}_c\cV^{n}{}_cR_{mn}\\
 \nonumber =&e^{-2\sigma}\big(-2\partial_\mu\partial^\mu\sigma-2\partial_\mu\partial^\mu\text{ln}\rho-\frac{3}{2}\partial_\mu\text{ln}\rho \partial^\mu\text{ln}\rho-P^\mu{}_{ab}P_\mu{}^{ab}\big).
\end{align}
The Einstein tensor enjoys a block diagonal structure, decomposing into external
\begin{align}
(G_{\text{Einstein}})_{\mu\nu}=&-\partial_\nu\partial_\mu \text{ln}\rho +2\partial_{(\mu} \text{ln}\rho\partial_{\nu)}\sigma -\frac{1}{2}\partial_\mu \text{ln}\rho\;\partial_\nu \text{ln}\rho-P_\mu{}^{ab}P_\nu{}_{ab}\\
&\hspace{2.5cm}+\eta_{\mu\nu}\big(\partial_\lambda\partial^\lambda\text{ln}\rho+\frac{3}{4}\partial_\lambda\text{ln}\rho \partial^\lambda\text{ln}\rho-\partial_\lambda\text{ln}\rho\partial^\lambda\sigma+\frac{1}{2}P^\lambda{}_{ab}P_\lambda{}^{ab}\big)\, ,
\notag
\end{align}
and internal components
\begin{align}
 \nonumber (G_{\text{Einstein}}   )_{mn}\!=&\,e^{-2\sigma}\rho\Big[\!-\cV_m{}^a\cV_n{}^b(\partial_\mu P^\mu{}_{ab}+Q_\mu{}^c{}_aP^\mu_{bc}+Q_\mu{}^c{}_bP^\mu_{ac})-\cV_m{}^a\cV_n{}^b\partial_\mu\text{ln}\rho P^\mu{}_{ab}\\[5pt]
 \nonumber&\hspace{1.2cm}+\cV_m{}^a\cV_n{}^a\big(\partial^\lambda\partial_\lambda\sigma+\frac{1}{2}\partial^\lambda\partial_\lambda\text{ln}\rho+\frac{1}{4}\partial_\lambda\text{ln}\rho\partial^\lambda\text{ln}\rho+\frac{1}{2}P_\lambda{}^{ab}P^\lambda{}_{ab}\big)\!\Big]\\[5pt]
 =&e^{-2\sigma}\rho\Big[\!-\rho^{-1}\cV_m{}^a\cV_n{}^b D_\mu(\rho P^\mu{}_{ab})\\
 &\hspace{1.1cm}+\cV_m{}^a\cV_n{}^a\big(\partial^\lambda\partial_\lambda\sigma+\frac{1}{2}\partial^\lambda\partial_\lambda\text{ln}\rho+\frac{1}{4}\partial_\lambda\text{ln}\rho\partial^\lambda\text{ln}\rho+\frac{1}{2}P_\lambda{}^{ab}P^\lambda{}_{ab}\big)\!\Big]\, ,
 \notag 
\end{align}
where we have introduced the SO(2) covariant derivative $D_\mu=\partial_\mu+Q_\mu$.
\subsection{Landau-Lifshitz pesudo-tensor}
The Landau-Lifshitz pseudo-tensor is defined as \cite{Landau:1975pou}
\begin{equation}
  t^{MN}_{LL}=-(G_{\text{Einstein}})^{MN}+\frac{1}{2}E^{-2}\partial_\alpha\partial_\beta\Big(E^2\big(G^{MN}G^{\alpha\beta}-G^{M\alpha}G^{N\beta}\big)\Big)\, .  
\end{equation}
First, it is straightforward to assess that in our Kaluza-Klein setup, it has a block diagonal structure. We opt for rewriting its correspondent blocks with indices lowered by the metric, namely
\begin{align}
    (t_{LL})_{\mu\nu}&=-(G_{\text{Einstein}})_{\mu\nu}+\frac{1}{2}\rho^{-2}\partial_\alpha\partial_\beta\Big(\rho^2\big(\eta_{\mu\nu}\eta^{\alpha\beta}-\delta_{\mu}^{\alpha}\delta_\nu^\beta\big)\Big)\\[5pt]
  (t_{LL})_{mn}&=-(G_{\text{Einstein}})_{mn}+\frac{1}{2}e^{-4\sigma}\cV_m{}^b\cV_p{}^b\cV_n{}^c\cV_q{}^c\partial_\alpha\partial^\alpha\Big(e^{2\sigma}\rho\cV^p{}_a\cV^q{}_a\Big)  \, .
\end{align}
We compute
\begin{equation}
\begin{aligned}
\frac{1}{2}\rho^{-2}&\partial_\alpha\partial_\beta\Big(\rho^2\big(\eta_{\mu\nu}\eta^{\alpha\beta}-\delta_{\mu}^{\alpha}\delta_\nu^\beta\big)\Big)
\\ &=+2\eta_{\mu\nu}\partial_\alpha\text{ln}\rho\partial^\alpha\text{ln}\rho-2\partial_\mu\text{ln}\rho\partial_\nu\text{ln}\rho +\eta_{\mu\nu}\partial_\alpha\partial^\alpha\text{ln}\rho-\partial_\mu\partial_\nu\text{ln}\rho\, ,
\end{aligned}
\end{equation}
and
\begin{align}
\nonumber\,\,\frac{1}{2}&e^{-4\sigma}\cV_m{}^b\cV_p{}^b\cV_n{}^c\cV_q{}^c\partial_\alpha\partial^\alpha\Big(e^{2\sigma}\rho\cV^p{}_a\cV^q{}_a\Big)\\[3pt]
\nonumber=\rho& e^{-2\sigma}\Big[\!\big(2\partial_\alpha\sigma\partial^\alpha\sigma+2\partial_\alpha\sigma\partial^\alpha\text{ln}\rho+\partial_\alpha\partial^\alpha\sigma+\frac{1}{2}\partial^\alpha\partial_\alpha\text{ln}\rho+\frac{1}{2}\partial^\alpha\text{ln}\rho\partial_\alpha\text{ln}\rho\big)\cV_m{}^a\cV_n{}^a\\
&\hspace{2.9cm} +(\cV_n{}^f\cV_m{}^a+\cV_n{}^a\cV_n{}^f)\mathcal{J}_{\mu\; e}{}^f P^{\mu\;ea}-\cV_n{}^e\cV_m{}^a\partial_\mu P^\mu{}_{ea}\Big].
\end{align}
In the end
\begin{align}\label{eq:LLpseudotensorext}
  (t_{LL})_{\mu\nu}&= -2\partial_{(\mu} \text{ln}\rho\partial_{\nu)}\sigma -\frac{3}{2}\partial_\mu \text{ln}\rho\;\partial_\nu \text{ln}\rho+P_\mu{}^{ab}P_{\nu\;ab}\\
 &\quad\, +\eta_{\mu\nu}\big(+\partial_\lambda\text{ln}\rho\partial^\lambda\sigma+\frac{5}{4}\partial_\alpha\text{ln}\rho\partial^\alpha\text{ln}\rho-\frac{1}{2}P^{\alpha\;ab}P_{\alpha\;ab}\big)\, ,
 \notag \\[5pt]
(t_{LL})_{mn}&=e^{-2\sigma}\rho\Big[+\rho^{-1}\cV_{m}{}^a\cV_{n}{}^b\partial_\mu\rho P^\mu{}_{ab}+2\cV_m{}^a\cV_n{}^b\Big( P_{\mu\;ac}P^{\mu\;c}{}_b\Big) \label{eq:LLpseudotensorint}
\\ 
&\hspace{1.75cm}+ \cV_m{}^d\cV_n{}^d\big(2\partial_\mu\sigma\partial^\mu\sigma+2\partial_\mu\sigma\partial^\mu\text{ln}\rho+\frac{1}{4}\partial_\mu\text{ln}\rho\partial^\mu\text{ln}\rho-\frac{1}{2}P^{\mu\;ab}P_{\mu\;ab}\big)\Big] \, .\notag 
\end{align}

A possible strategy to adopt in looking for the uplift of $\nu_2$ is to look at its linearised deformation, which in $D=2$ language can be always written as \eqref{eq:nu} expressed in terms of physical fields (at linear order in the equations of motion),
\begin{equation}\label{eq:lineariseddef}
  {\nu_2}_{\text{lin}}=(2\eta^{\alpha\beta}\eta^{\gamma\sigma}-\eta^{\alpha\gamma}\eta^{\beta\sigma})\left(\partial_{(\alpha}\sigma\partial_{\gamma)}\rho-\frac{\rho}{2}\text{Tr}(P_\alpha P_\gamma)\right)\left(\partial_{(\beta}\sigma\partial_{\sigma)}\rho-\frac{\rho}{2}\text{Tr}(P_\beta P_\sigma)\right)\, .  
\end{equation}
Then, one immediate guess for the uplift of the latter operator would be the Landau-Lifshitz pseudo-tensor, \eqref{eq:LLpseudotensorext}, \eqref{eq:LLpseudotensorint}, contracted as
\begin{equation}\label{eq:uplift1?}
    \mathcal{O}_{LL} = (G^{MP}G^{NQ}-\frac{1}{2}G^{MN}G^{PQ}) (t_{LL})_{MN}(t_{LL})_{PQ}\, ,
\end{equation}
or in a more customary way, as prescribed, for example in \cite{Taylor:2018xcy} (eq. (1.3))
\begin{equation}\label{eq:uplift2?}
    \mathcal{O}_{LL} = (G^{MP}G^{NQ}-\frac{1}{3}G^{MN}G^{PQ}) (t_{LL})_{MN}(t_{LL})_{PQ}\, ,
\end{equation}
where $G_{MN}$ is the $D=4$ metric obtained from \eqref{eq:vierbein1} as $G_{MN}=E_M{}^A\eta_{AB}E_N{}^B$. Unfortunately, upon inserting the KK-reduction ansatz \eqref{eq:vierbein1}, neither  \eqref{eq:uplift1?} nor \eqref{eq:uplift2?} matches \eqref{eq:lineariseddef}: this can be inferred immediately by the presence (after having appropriately rescaled $\sigma$) of quartic derivatives terms of the kind
\begin{equation}
 \partial_\alpha \rho \partial^\alpha\rho\partial_\beta\rho\partial^\beta\rho\,, \quad   \partial_\alpha \rho \partial_\beta\rho P^{\alpha\;ab} P^{\beta}{}_{ab},  
\end{equation}
which do not appear in \eqref{eq:lineariseddef}.  We have computed explicitly \eqref{eq:uplift1?} and \eqref{eq:uplift2?} with the help of the software Cadabra, but have chosen to refrain from displaying the full result, which takes up several pages and which we do not deem particularly informative.
%
%
%
%
\bibliographystyle{JHEP}
\bibliography{Bib}
\end{document}